\documentclass{aa}
\usepackage{graphicx}
\usepackage{gensymb}
\usepackage{multirow}
\usepackage{amsmath}

\usepackage{txfonts}
\usepackage{xcolor}
\definecolor{purple}{rgb}{0.5,0,0.87}

\begin{document} 

\title{Galactic cirri in deep optical imaging}
\authorrunning{Román, Trujillo and Montes}

%\subtitle{I. Overviewing the $\kappa$-mechanism}

   \author{Javier Román\inst{1}\inst{2}\inst{3}\thanks{   \email{jromanastro@gmail.com}}
   \and
   Ignacio Trujillo\inst{1}\inst{2}
   \and
   Mireia Montes\inst{4}
   }

   \institute{Instituto de Astrof\'{\i}sica de Canarias, c/ V\'{\i}a L\'actea s/n, E-38205, La Laguna, Tenerife, Spain
   \and Departamento de Astrof\'{\i}sica, Universidad de La Laguna, E-38206, La Laguna, Tenerife, Spain
   \and Instituto de Astrof\'isica de Andaluc\'ia (CSIC), Glorieta de la Astronom\'ia, 18008 Granada, Spain
   \and School of Physics, University of New South Wales, Sydney, NSW 2052, Australia
   }
   \date{\today}

% \abstract{}{}{}{}{} 
% 5 {} token are mandatory
 
  \abstract
{The ubiquitous presence of Galactic cirri in deep optical images represents a major obstacle to study the low surface brightness features of extragalactic sources. To address this issue, we have explored the optical properties of cirri using \textit{g}, \textit{r}, \textit{i}, and \textit{z} bands in the Sloan Digital Sky Survey (SDSS) Stripe82 region. Using state-of-the-art, custom made, image processing techniques, including the modeling and removal of the instrumental scattered light produced by the stars, we managed to isolate the optical diffuse emission by the cirri, allowing their photometric characterization. We find that their optical colors are driven by the dust column density: The cirri become redder as their 100 $\mu$m emission increases. Remarkably, the optical colors of the Galactic cirri differ significantly from those of extragalactic sources, with a characteristic bluer \textit{r-i} color for a given \textit{g-r}, allowing one to {detect} these by using a simple color relation. Our results show the high potential of deep multi-band optical photometry, on its own, identifying the presence of cirri at a higher spatial resolution than those provided by far-infrared observations. The combination of very deep data and multi-band photometry (as the one produced by LSST {and Euclid}) would make it possible to build dust maps of unprecedented quality.}

   \keywords{ ISM: clouds -- ISM: dust, extinction -- Techniques: image processing -- Techniques: photometric
               }

   \maketitle
%
%________________________________________________________________

\section{Introduction}

The current generation of deep optical surveys allows us, for the first time, to unravel critical issues of the hierarchical evolution of galaxies that are intrinsically connected with the low surface brightness Universe. For instance, the detection of galaxies of decreasing surface brightness \citep[e.g.,][]{1984AJ.....89..919S, 1988ApJ...330..634I, 1997AJ....114..635D, 2005ApJ...631..208B,2015ApJ...809L..21M,2016A&A...588A..89J, 2018ApJ...863L...7M}, the stellar halos and tidal features surrounding nearby galaxies \citep[e.g.,][]{2005ApJ...635..931B, 2006MNRAS.365..747A,2008ApJ...689..936J, 2010AJ....140..962M, 2015MNRAS.446..120D, 2016ApJ...823..123T} or the intracluster light in galaxy clusters \citep[e.g.,][]{1991ApJ...369...46U, 2005ApJ...631L..41M, 2010ApJ...720..569R, 2014ApJ...781...24G, 2014ApJ...794..137M, 2018MNRAS.474..917M} are all crucial observational pillars for testing the current $\Lambda$-Cold Dark Matter ($\Lambda$CDM) cosmological paradigm \citep[e.g.,][]{1999ApJ...524L..19M, 1999ApJ...522...82K, 2010MNRAS.406..744C, 2015MNRAS.451.2703C}.

The study of the faintest regions of the galaxies through the star counting technique reaches a depth equivalent to 30 mag arcsec$^{-2}$ when detecting galaxy satellites \citep[][and references therein]{2012AJ....144....4M} and 32 mag arcsec$^{-2}$ when exploring galactic substructures as tidal features and stellar streams \citep[e.g.,][]{2009MNRAS.395..126I, 2011ApJ...738..150T, 2014ApJ...780..128I, 2014ApJ...787...19M, 2018ApJ...868...55M}. However, the resolved star count method is only able to study the Local Universe, as it is limited by the spatial resolution of the images \citep[e.g., 16 Mpc using the Hubble Space Telescope;][]{2012MNRAS.421..190Z} and stochasticity \citep{2010AdAst2010E..21W}. For this reason, integrated photometry studies are the only way to explore the low surface brightness components of galaxies beyond the Local Group. However, the use of integrated photometry is affected by systematic effects, severely restricting the surface brightness limits of the optical data. In particular, we can list the following: i) poor data reduction processes that can create artificial gradients and different artifacts in the images; ii) incorrect subtraction of the sky background that removes or over-subtracts the low surface brightness information around extended sources; and iii) {instrumental} scattered light contamination from sources due to the extended point spread function (PSF) and internal reflections present in the images. Different approaches have been used to try to minimize these effects, such as better astronomical data reduction pipelines \citep[][]{2015ApJS..220....1A, 2019A&A...621A.133B}, improvements in the observational techniques \citep[e.g.,][]{2012ApJS..200....4F, 2015MNRAS.446..120D, 2016ApJ...823..123T}, a better characterization of the PSF and internal reflections due to optical instrumentation \citep[][]{2009PASP..121.1267S,2014A&A...567A..97S, Infante}, or the development of simple optic telescopes with the goal of minimizing the presence of instrumental scattered light \citep[][]{2014PASP..126...55A,2017ApOpt..56.8639M,2017IAUS..321..199V,2019MNRAS.490.1539R}.

Despite all the observational and technical advances aiming to improve the quality of deep optical datasets, the presence of interstellar material from our own Galaxy reflecting starlight \citep[][]{1937ApJ....85..213E, 1941ApJ....93...70H, 1976AJ.....81..954S, 1979A&A....78..253M} is unavoidable, thus harming the study of the faintest regions of galaxies. These filamentary clouds of dust, called "Galactic cirri", have their peak emission in the far-infrared due to their low temperature \citep[][]{1984ApJ...278L..19L, 2010ApJ...713..959V}. However, Galactic cirri are detectable in optical { \citep[][]{1985A&A...145L...7D, 1987A&A...184..269L, 2008ApJ...679..497W,2016A&A...593A...4M}} and ultraviolet \citep[][]{1997ApJ...481..809W, 2015A&A...579A..29B} wavelengths, {reflecting starlight in those spectral ranges}. With the increasing depth of optical observations, the presence of cirri has become a major obstacle in the study of faint extragalactic features even at high Galactic latitudes, mimicking their shape and brightness, thus creating confusion \cite[e.g.,][]{2009AJ....137.3009C, 2010MNRAS.403L..26C, 2010A&A...516A..83S, 2010ApJ...720..569R, 2010MNRAS.409..102D, 2013AJ....146..126C, 2014ApJ...789..131H, 2016ApJ...825...20B, 2018MNRAS.475L..40D, 2018A&A...616A..42B, 2018A&A...619A.163R}. Far infrared or submillimeter full-sky maps, as the IR Astronomical Satellite (IRAS) mission or the Planck Space Observatory, are used to identify the presence of cirri in the optical images. However, the poor spatial resolution of the far-infrared and submillimeter images of these instruments \citep[FWHM $\approx$ 5 arcmins;][]{2005ApJS..157..302M, 2003NewAR..47.1017L} makes them inefficient for the study of small angular scale features. Only with the availability of far infrared data of higher spatial resolution, as the ESA Herschel Space Observatory \citep[FWHM of 18 arcseconds in the 250 $\mu$m band;][]{2010A&A...518L...1P} it has been possible to identify correctly, or even remove, the dust foreground contamination \citep[see][]{2017ApJ...834...16M}. Unfortunately, the Herschel Space Observatory data only covers a modest portion of the sky. Therefore, it would be desirable to use the optical data itself to identify the presence of cirri without the need to resort to complementary data at other wavelengths. This means, ultimately, to define accurately the range in optical colors of the Galactic cirri, allowing to discern between them and extragalactic sources, if possible. 

Previous studies of the optical properties of the Galactic cirri {\citep[e.g.,][]{1968ApJ...152...59W,1970A&A.....8..273M,1989ApJ...346..773G, 2008ApJ...679..497W, 2013ApJ...767...80I}} have neglected important systematic effects, such as the presence of {instrumental} scattered light by bright stars due to the PSF {at large radii}. This effect, for instance, can have a significant impact on the photometry of the cirri due to their extremely low surface brightness, especially in areas where the dust clouds have a low column density or surface brightness.

The imminent arrival of the {Legacy Survey of Space and Time\footnote{Carried out at the Vera Rubin Observatory.}} \citep[LSST; ][]{2009arXiv0912.0201L, Brough2020} the Euclid mission \citep{2011arXiv1110.3193L} and the new generation of extremely large telescopes, will light up the presence of Galactic cirri at any Galactic latitude. It is, therefore, a priority for the extragalactic community to address the problem of the cirri contamination in order to study the low surface brightness Universe. {Addressing this contamination issue} is also crucial in other cases, {such as} the characterization of the Cosmic Microwave Background and its polarization \citep[e.g.,][]{2011A&A...536A..19P, 2014JCAP...04..014D} or the Infrared Cosmic Background \citep[e.g.,][]{2013ApJ...768...58T}. In this work, we aim to carry out a comprehensive photometric analysis of the Galactic cirri, taking into account the possible systematic uncertainties derived from the extremely low surface brightness regime. The goal is to provide the colors of the cirri in optical wavelengths {in order to discern these from extragalactic sources}. 

This paper is structured as follows: In Section 2 we describe the data and its processing. The correlation between the far IR and optical bands is explored in Section 3. In Section 4, we obtain the optical colors of the Galactic cirri. We address our main conclusions in Section 5. We use the AB photometric system throughout this work.

\section{Data and data processing}\label{sec:proccesing}

A common problem affecting deep optical surveys is the over-subtraction of flux around extended sources. This is the result of an aggressive sky subtraction. A full discussion about this problem and a strategy to solve it can be found in \citet{2019A&A...621A.133B} using the Hubble Ultra-Deep Field dataset. {Early} public versions of the Hyper Suprime-Cam Subaru Strategic Program (HSC-SSP) \citep{2018PASJ...70S...4A} or the Dark Energy Camera Legacy Survey (DECaLS) \citep{2016MNRAS.460.1270D} are severely affected by an aggressive sky subtraction. This circumstance prevents the study of extended diffuse emission, which misidentified as the sky itself, it is systematically subtracted. To overcome this problem, we use data from the IAC Stripe82 Legacy Survey \citep{2016MNRAS.456.1359F}. This dataset consists in a new reduction of the SDSS Stripe82 data \citep[][]{2009ApJS..182..543A} with the aim of preserving the faintest extended features, minimizing the over-subtraction around sources. The Stripe82 region covers a 2.5 degree wide stripe along the Celestial Equator in the Southern Galactic Cap (-50\degree $<$ R.A. $<$ 60\degree, -1.25\degree $<$ Dec.$<$ 1.25\degree) reaching a total of 275 square degrees. The total exposure time per field is around 1 hour, using the 2.5m Telescope at Apache Point Observatory (SDSS) in all the five SDSS filters \textit{(u, g, r, i, z)}. The pixel scale is 0.396 arcsec and the average seeing is around 1 arcsec. In this work, we use the new set of rectified images of the survey described by \cite{2018RNAAS...2c.144R} in which residuals of the co-adding process have been removed. This produces high-quality and homogenized deep images, allowing a clean photometry of the faintest structures. The average surface brightness limits of the dataset are $\mu_{lim}$ (3$\sigma$; 10\arcsec$\times$10\arcsec) = 28.0, 29.1, 28.6, 28.2 and 26.6 mag arcsec$^{-2}$ for the \textit{u, g, r, i} and \textit{z} bands respectively. We describe in detail how these limits are calculated in Appendix \ref{sec:Depth}.

Although the IAC Stripe82 dataset preserves the diffuse emission, the photometry of extended sources is not straight forward to obtain. For instance, the {instrumental} scattered light produced by the PSF wings of the stars strongly contaminates the faint cirri emission. Additionally, {a comprehensive} masking of sources is necessary to obtain reliable photometry of the cirri. In the following, we describe the data processing performed in order to isolate the flux of the Galactic cirri.

\subsection{Modeling the {instrumental} scattered light produced by the stars}

The removal of the {instrumental} scattered light produced by stars is becoming a common technique to obtain clean photometry of extremely low surface brightness sources. For instance, \cite{2009PASP..121.1267S} modeled the internal reflections produced by stars observed with the Burrell Schmidt telescope \citep[{and used in }][]{2010ApJ...720..569R, 2014ApJ...791...38W, 2015ApJ...800L...3W, 2016ApJ...826...59W, 2017ApJ...851...51W, 2017ApJ...834...16M, 2018ApJ...858L..16W}, \cite{2016ApJ...823..123T} modeled the {instrumental} scattered light by stars in very deep observations of the UGC 00180 galaxy using a PSF characterization of the 10.4m GTC telescope and \citet[][]{2017A&A...601A..86K} modeled the internal reflections and PSF using images of the Canada France Hawaii Telescope (CFHT) to provide reliable deep photometry of galaxies. {Other efforts, such as detection of image artifacts in the images \citep{2016A&C....16...67D}, have also been performed}.

The specific characteristics of the internal reflections due to the optical instrumentation of the telescope is an important issue to take into account when modeling the stars. Depending on the extension of the reflections, the sum of all of them can contaminate a significant fraction of the total area of the image. This circumstance requires a modeling of the reflections. In the specific case of the IAC Stripe82 Legacy Survey, the internal reflections are located at the same spatial position in each single exposure, and therefore, in the final coadd. Additionally, these reflections only subtend a small angular size \citep[up to 30'', see][]{Infante}, being relatively faint and only noticeable for bright stars. Hence, it is not necessary to model these reflections, but only the PSF (image of a point source through the main optical path, without reflections on the CCD). The sum of the {instrumental} scattered light by all the stars in a given field due to the PSF is what we call the "{instrumental} scattered light field".

Obtaining this {instrumental} scattered light field means characterizing accurately the PSF. Once the PSF model is obtained, the {instrumental} scattered light field is defined by the position and fluxes of the stars in the image. {Thus the only information needed are the coordinates and fluxes of the stars to be modeled.} In Appendix \ref{sec:PSF} we describe the process followed to model the PSF for the IAC Stripe82 Legacy Survey in the five SDSS bands. 

To obtain the exact coordinates and fluxes of the stars, there is a number of practical problems, like the inner saturated region of the bright stars or the presence of adjacent or overlapping sources. These difficulties make the process of fitting stars not straightforward. In this work, we have created a specific pipeline that obtains precise positions and accurate photometry of stars, even if they are saturated. The main tasks of this pipeline are summarized on what follows.

\begin{figure}
  \centering
   \includegraphics[width=\columnwidth]{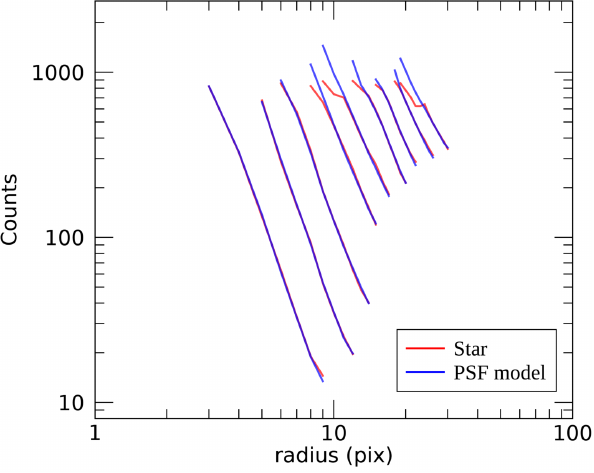}%0.635
    \caption{Surface brightness profiles in the \textit{g} band of real stars together with the fitted PSF models in a random field of the Stripe82 region. The plot shows the profiles of the fitted stars (red) and the profile of the flux scaled PSF models (blue). Nine stars are fitted in this example, with magnitudes ranging approximately between 7.5 and 15.5 in the \textit{g} band. The profiles cover the radial region used to calculate the flux of each star. The radial region selected for the profile fitting is dynamic for each star and depends on the circumstances of saturation, crowding or magnitude of the star (see main text). We note that the saturation region appears above approximately 800 counts, creating severe deviations from the PSF model in the inner region of the profiles of the brightest stars due to CCD bleeding.}

   \label{fig:profiles}
\end{figure}

\begin{figure}
  \centering
   \includegraphics[width=\columnwidth]{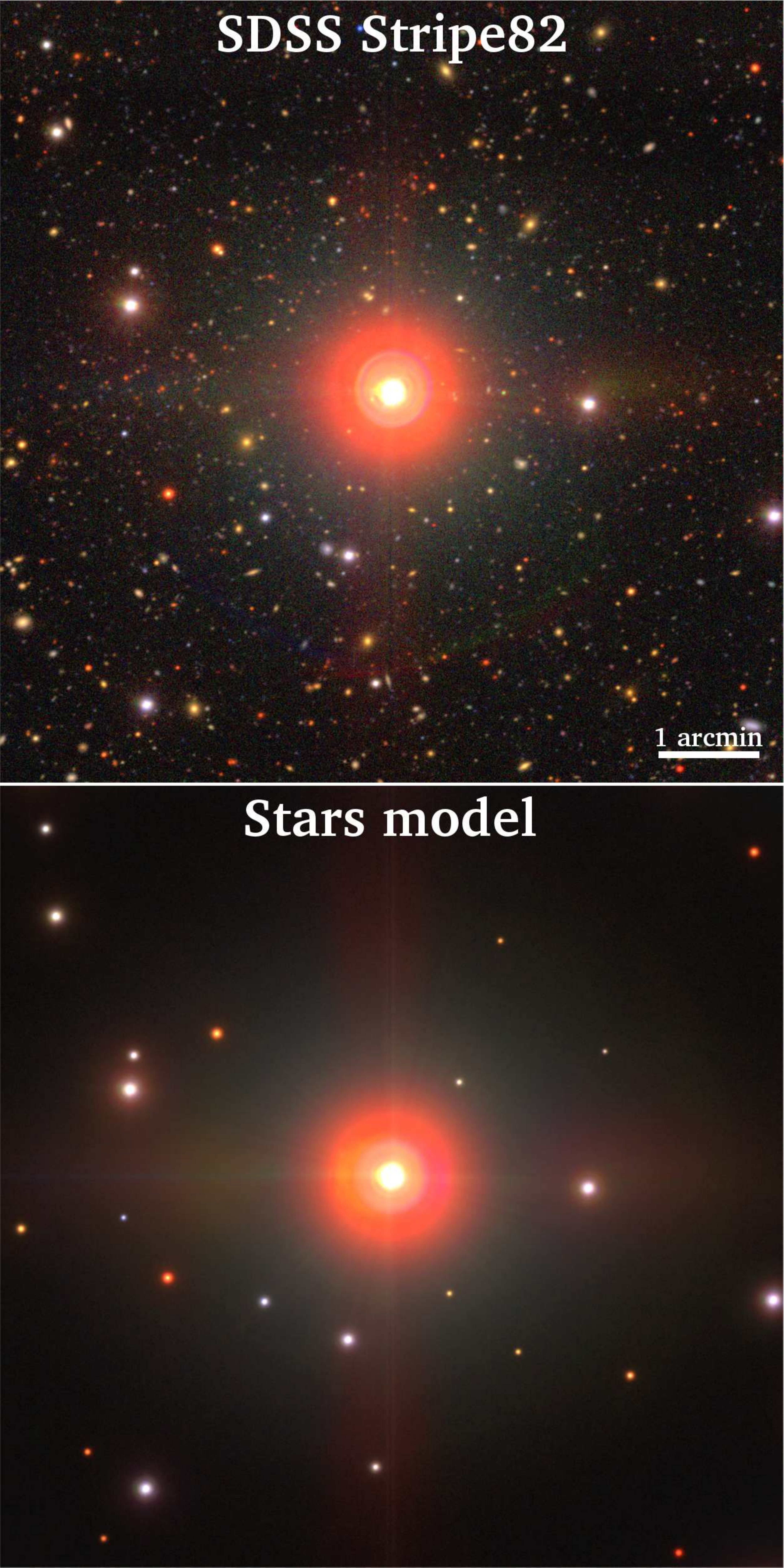}%0.635
    \caption{Color composite image of a real bright star and its surrounding field in the Stripe82 region using \textit{g}, \textit{r} and \textit{i} SDSS bands (upper panel). In the bottom panel we show a model of the bright and other fainter stars of the same field.}

   \label{fig:PSF_model}
\end{figure}

For each {of the fields studied here} we produced a catalog of stars. {To do that},  we ran \texttt{SExtractor} \citep{1996A&AS..117..393B} obtaining the position and stellarity for each source in the field. We classify as stars those sources with stellarity (\texttt{CLASS\_STAR}) higher than 0.75 in the \textit{r-deep} band, a deep \textit{g} + \textit{r} + \textit{i} combined image provided by the IAC Stripe82 survey. This filters stars by exclusively selecting PSF-like sources. However, very bright stars do not match this criterion due to the strong inner saturation and bleeding {producing lower values of stellarity}. Since these stars are very bright, therefore very few, we included their coordinates manually to complete the final catalog. 

The next step is to obtain accurate spatial coordinates for the selected stars, which is crucial to model the PSF \citep{2018AJ....156...14L}. For each star, we masked the saturated region (if present) by masking pixels above the saturation threshold of each band. We also masked the sources that are adjacent, or overlapping, to the stars using the segmentation maps provided by \texttt{SExtractor}. Because these detection maps do not cover well enough the sources, in particular the faint external regions, we expanded the masks to cover most of the flux of the sources {by smoothing that mask with a gaussian kernel of 2 pixels}. To obtain accurate spatial coordinates for each star, we fit a Moffat function to the masked star using the \texttt{IMFIT} \citep{2015ApJ...799..226E} package, {obtaining the coordinates of the center from the fit}. 

Once the star is masked and centered, we performed the flux fitting. {To do that, we derived the radial profiles of both the PSF and the star, then, we scaled the radial profile of the PSF to match in flux the profile of the star}. We adopted this methodology as profiles are more resistant to masking residuals, crowding or any artifact that could remain in the images. For each star, we selected a range in radius for the scaling. This range is dynamic and depends on the tentative magnitude provided by \texttt{SExtractor} (imprecise due to saturation among other factors) and the crowding of adjacent sources to the star. In general terms, the higher the luminosity of the star, the larger the range in radius, and if the star has {a bright adjacent source (avoiding the masking in the more external regions)}, this range will be truncated to the point where the profile increases with the radius. With this procedure, we select a range in the profile with a high signal to noise, which depends on the luminosity of the star, avoiding reaching areas where adjacent sources could deviate the flux measurement. Once the profile of the star is obtained, we scaled it to match the normalized PSF profile, giving the flux of the star. To obtain the most accurate measurement possible, we averaged the flux ratios between the star and the PSF profiles at each radius, discarding outliers beyond 2$\sigma$. This provides a robust flux value for the star. Fig. 1 shows an illustrative example of {the accuracy of this methodology for nine star profiles of different magnitudes}.

Repeating this process for the catalog of stars, we obtained the positions and integrated fluxes of all the stars in the field. For this work, we modeled all the stars up to magnitude 15 in the r-band for computational efficiency. We found that for a typical Stripe82 field, the {instrumental} scattered light field produced by stars fainter than 15 magnitude is negligible.

Finally, we use the software SWARP \citep{2002ASPC..281..228B} to create the {instrumental} scattered light field by placing PSFs with the calculated positions and fluxes of the stars over a blank field. We show in Fig. \ref{fig:PSF_model} an example of an arbitrary field modeled with our procedure. The average number of fitted stars is approximately 400 per square degree. Due to the strong presence of artifacts, that is to say "ghosts", in the \textit{u} band, and the high sky background fluctuations in this band, we restrict ourselves to the more reliable \textit{g}, \textit{r}, \textit{i} and \textit{z} bands for further analysis.

\subsection{Properties of the {instrumental} scattered light field produced by stars}

\begin{figure*}
  \centering
   \includegraphics[width=0.95\textwidth]{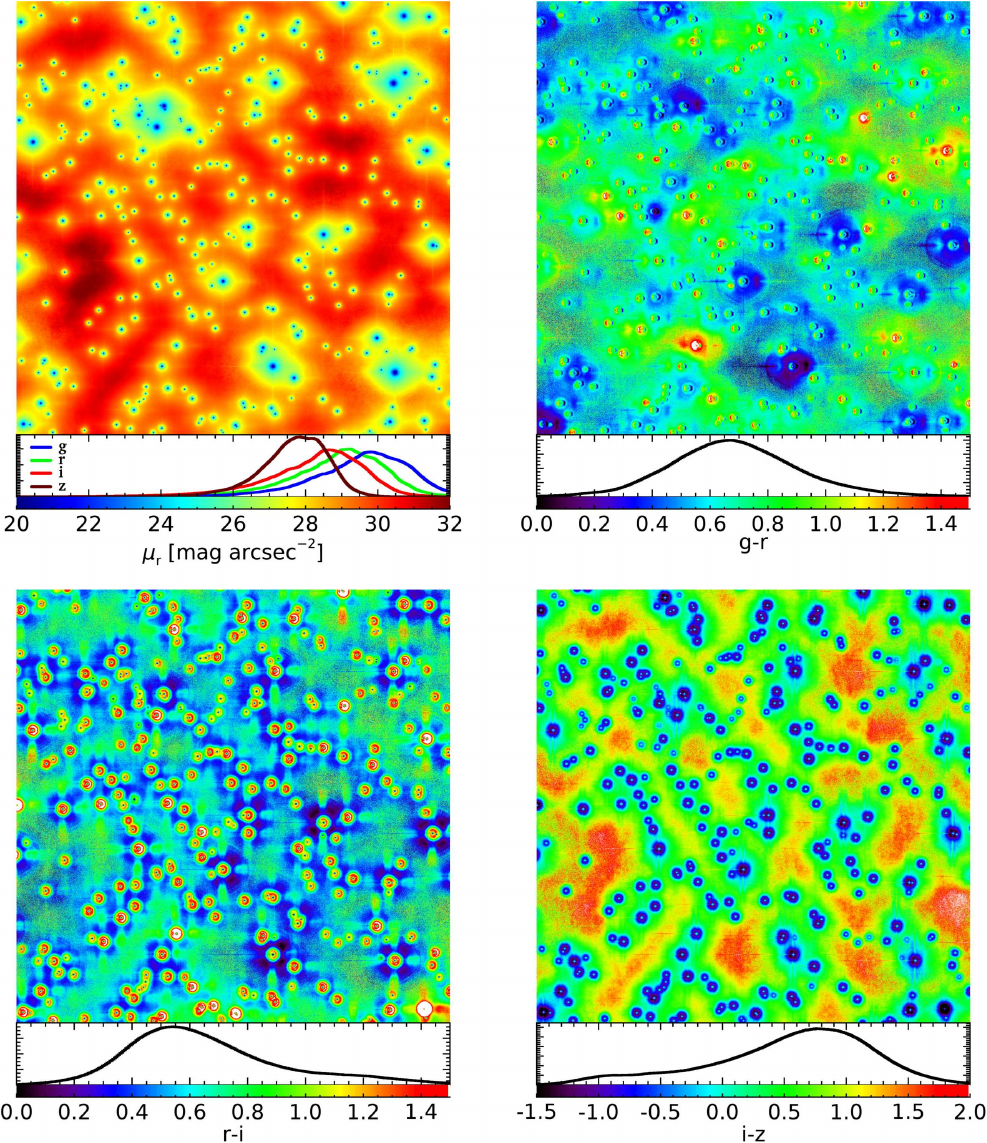}%0.635
    \caption{Photometric properties of the {instrumental} scattered light field produced by the stars on a random field of 1 $\times$ 1 degree of the IAC Stripe82 Legacy Survey. The upper left panel shows the surface brightness distribution of the pixels of the field in the \textit{r} band. The upper right, bottom left and bottom right panels show the \textit{g-r}, \textit{r-i} and \textit{i-z} colors respectively. The bottom plots in each panel represent the distribution histograms.}

   \label{fig:SL_colors}
\end{figure*}

Obtaining the {instrumental} scattered light field model provides interesting information about its effect on the data. The availability of the \textit{g}, \textit{r}, \textit{i} and \textit{z} SDSS bands gives us the possibility to obtain {a photometric description of the colors of the} {instrumental} scattered light produced by the stars. In Fig. \ref{fig:SL_colors}, we show a random field of $1\degree\times1\degree$ of the IAC Stripe82 Legacy Survey. In the upper left panel, we present the surface brightness map produced by the stars in the \textit{r} band. In the lower plot of that panel, we show the histograms of surface brightness for all the pixels in the image for the \textit{g}, \textit{r}, \textit{i} and \textit{z} bands. As can be seen, the surface brightness distributions of the {instrumental} scattered light field vary between different bands. In the case of the \textit{r} band, there is an average surface brightness of approximately 29 mag arcsec$^{-2}$, similar to the value reported by \cite{2016ApJ...823..123T} (using the GTC telescope) in that same band \citep[see also work by][]{2009PASP..121.1267S}. The average surface brightness of the {instrumental} scattered light field increases as the band gets redder, ranging from a typical surface brightness of 30 mag arcsec$^{-2}$ in the \textit{g} band to 28 mag arcsec$^{-2}$ in the \textit{z} band. It is of course, expected, as the average star is brighter in redder optical bands. Another way to look at this is shown in the other panels of Fig. \ref{fig:SL_colors}, where the color maps of the {instrumental} scattered light field are shown, together with their color distributions. {The three color maps shown ($g-r$, $r-i$ and $i-z$) have mean red colors}. Remarkably, the colors of the {instrumental} scattered light are far from having an homogeneous distribution in the image, suffering from strong color variations at small spatial scales. This effect is particularly striking in those colors containing the \textit{i} band. The reason is the complex PSF structure in this band, with a bump within 30 arcsec from the center \citep[][see also Fig.  \ref{fig:PSFs_profiles}]{2008MNRAS.388.1521D,Infante}. This produces complex color patches around the stars, clearly noticeable in the brightest star in Fig. \ref{fig:PSF_model}.

\begin{table}
\centering
\caption{Summary of \texttt{SExtractor} configuration parameters applied for the different masking layers. The enlarged mask size is given in pixels units (0.396 arcsec).}
{\footnotesize
\label{tab:params}
\begin{tabular}{c  ccccc}
\hline
Mask & \texttt{DETECT} & \texttt{DETECT} & \texttt{DETECT} & \texttt{BACK} & Enlarged \\
 & \texttt{MINAREA} & \texttt{MAXAREA} & \texttt{THRESH} & \texttt{SIZE} & mask size \\
1 & 5 & 150 & 3 & 10 & 0  \\
2 & 150 & 500 & 3 & 10 & 5 \\
3 & 450 & 1500 & 2 & 10 & 15  \\
4 & 1900 & 7500 & 4 & 200 & 20  \\
5 & 1000 & 2500 & 3 & 10 & 30  \\
6 & 7000 & inf. & 4 & 100 & 30 \\
\hline
\end{tabular}
}
\end{table}

Although the vast majority of the {instrumental} scattered light field in an image has low surface brightness, it can have a significant impact on the photometry of extremely faint sources, as is the case of the Galactic cirri. In fact, the average surface brightness of the Galactic cirri analyzed in this work is well below 26 mag arcsec$^{-2}$ in the \textit{r} band. Therefore the {instrumental} scattered light field could strongly affect the global photometry of the cirri. We note that the results shown in this section are only descriptive of the data used in this work: SDSS data from the IAC Stripe82 Legacy Survey. The use of another dataset with different PSFs could create different {instrumental} scattered light fields and derived photometric properties.

\subsection{Isolating the extended diffuse emission of the Galactic cirri}

The modeling and subtraction of the stars produces images almost free of contamination by {instrumental} scattered light. However, faint stars, galaxies, residuals of the {instrumental} scattered light subtraction and different artifacts still remain in the image (see Fig. \ref{fig:Processing1}). In order to isolate the diffuse emission of the Galactic cirri in the images, it is necessary to mask accurately all the sources. However, the masking of low surface brightness structures is not an easy task. For example, \texttt{SExtractor} is not able to detect the faintest outer regions of the sources, producing an incomplete masking and a potential source of contamination. The reason is that \texttt{SExtractor} detects sources targeting exclusively stars and galaxies, being inefficient in the detection of amorphous structures, as in the case of cirri. However, this ineffectiveness in the detection of complex low surface brightness features works to our advantage, since only stars and galaxies are aimed to be masked. After testing several options, we ended up with a modified masking process based on SExtractor. It consists of six different layers of masking, focused in the detection of different sizes and deblending levels (sources over sources), each one with a different {widening of the mask}. In Table \ref{tab:params}, we list the \texttt{SExtractor} parameters for each masking layer. This masking sequence manages to cover all the sources, exposing only the diffuse emission by the cirri in the images. Since our goal is precision photometric measurements of Galactic cirri, any possible systematic uncertainty should be treated conservatively. {This means that the masks are systematically larger than the actual size of the sources. However, due to the large area occupied by the cirri in relation to the sources to be masked, that loss of area is acceptable for the purposes of this work. 

\begin{figure*}
  \centering
   \includegraphics[width=1.0\textwidth]{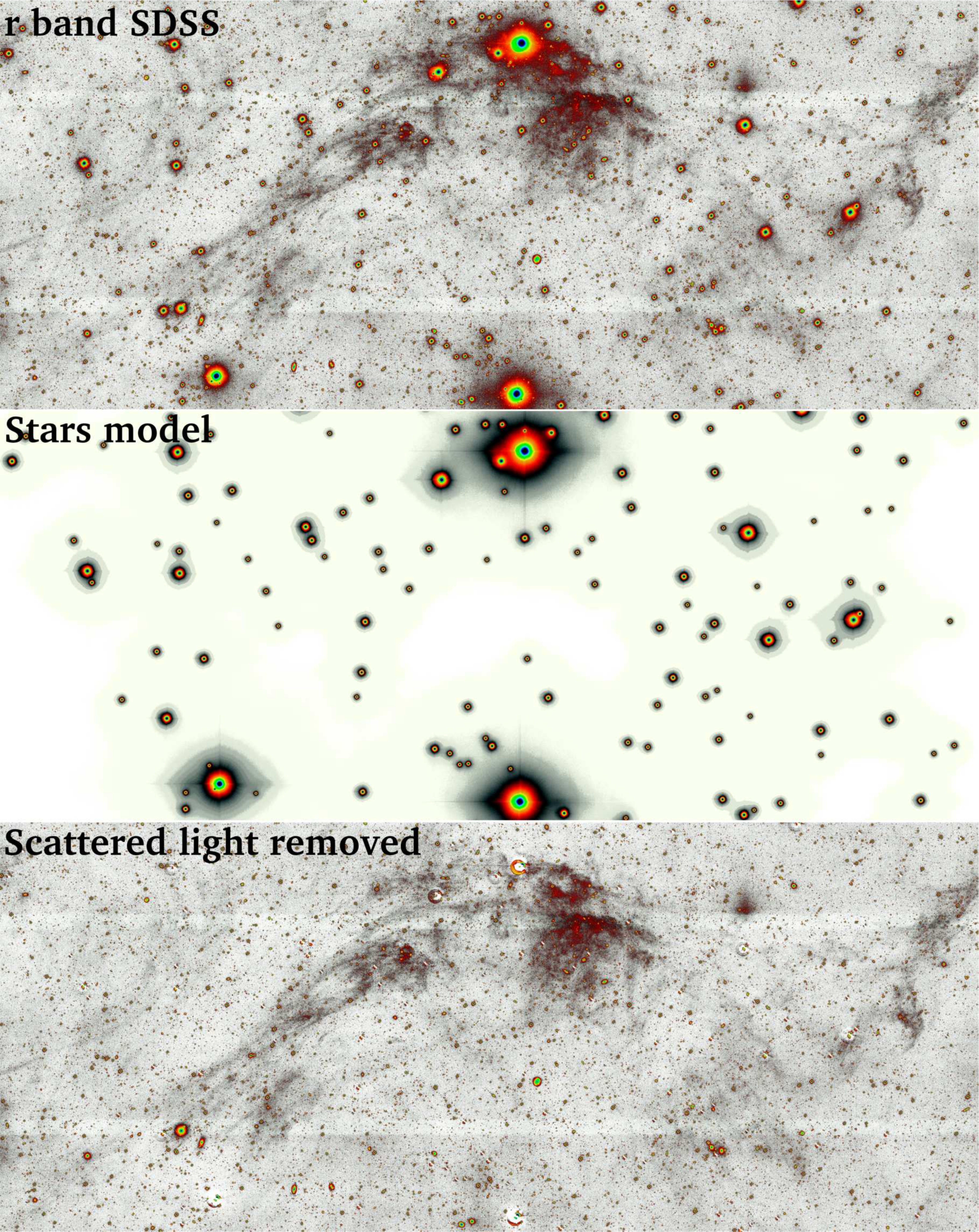}%0.635
    \caption{Example of the process of removing the {instrumental} scattered light produced by the stars in Stripe82 images. Top panel: Original IAC Stripe82 field in the r-band. Middle panel: Model of the {instrumental} scattered light field produced by the stars. Bottom panel: Same as the top panel but with the {instrumental} scattered light field subtracted. The pixel scale is 0.396 arcseconds.}

   \label{fig:Processing1}
\end{figure*}

\begin{figure*}
  \centering
   \includegraphics[width=1.0\textwidth]{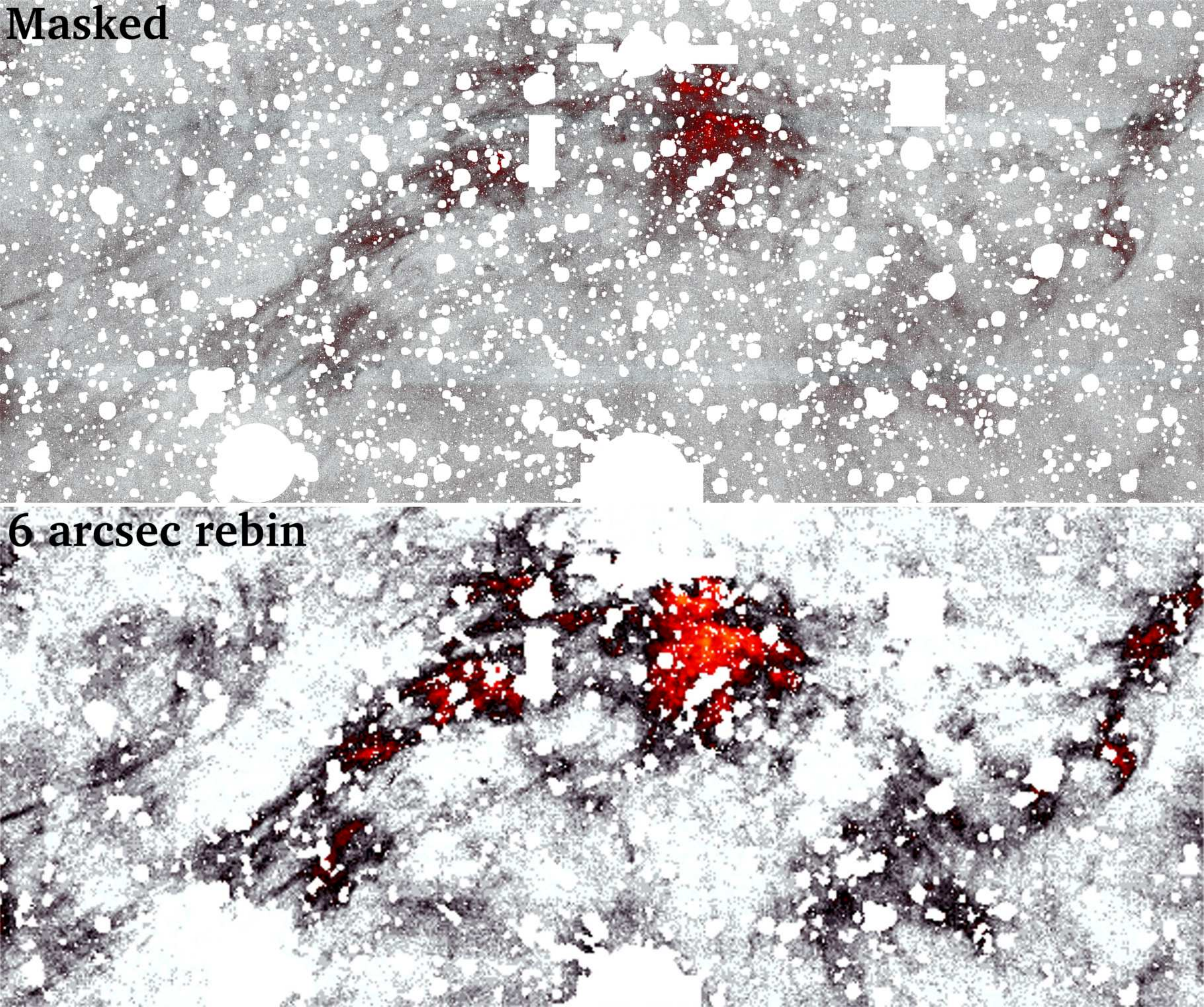}%0.635
    \caption{Same example as in Fig. \ref{fig:Processing1} showing the masking process (upper panel) and rebinning (bottom panel). The images {have the instrumental} scattered light field removed. Masked regions appear as white color.}

   \label{fig:Processing2}
\end{figure*}

The masking algorithm has as input the original \textit{r-deep} band image of the field, with all the stars present in the image. The masks obtained from the different steps of the algorithm described above are combined in a unique mask, which is applied to the images in the different bands corrected from the {instrumental} scattered light field. We note that the presence of stars in the input image facilitates the masking of the centers of the removed stars (only the centers, since \texttt{SExtractor} is not able to mask the PSF wings of the stars), covering all the subtraction residuals. The application of this masking process produces an excellent isolation of the diffuse emission of the cirri. However, "ghosts" and other artifacts still remain in the images. Nevertheless, these are well documented\footnote{\url{ http://skyserver.sdss.org/dr12/en/tools/places/page6.aspx}} and are easily identified visually in single band images or through color maps, presenting extreme colors against real sources. Therefore, we manually masked those artifacts. {In addition, we also mask any residual that could remain on the images, including off-field stars whose instrumental scattered light is present in the studied areas.} We show an illustrative example of the final result of the entire masking process in Fig. \ref{fig:Processing2}. 

As a final step, we increase the pixel size of the images to enhance the signal to noise ratio of the final image by reducing the Poisson noise as the result of averaging the original pixels into a larger one. Due to the large angular size of the Galactic cirri, there is a large margin to increase the pixel size without a reasonable loss of spatial resolution. We choose a  final pixel scale of 6 arcsec, as a compromise between optimal spatial resolution and the depth of the images. With this pixel size the average surface brightness limits in pixel to pixel variations are $\mu_{lim} (3\sigma;\, 6''\times6'')$~=~28.5, 28.0, 27.6 and 26.0 mag arcsec$^{-2}$ for the \textit{g}, \textit{r}, \textit{i} and \textit{z} bands respectively (see Appendix \ref{sec:Depth}). 

After rebinning we found a small fraction of pixels (less than 1 percent) whose values are much higher than adjacent pixels. These pixels are very prominent with respect to the soft background light of the diffuse emission. By inspecting the images, we detected that these hot pixels are the result of residuals in the masking process in the original pixel scale. These are typically faint tidal features or haloes of galaxies, not detectable in the processed images but enhanced in the rebinned ones. To remove these residuals, we carried out a last masking step in the rebinned images. Due to the isolated nature of these ``hot'' pixels in comparison with the smooth background of the diffuse emission, we masked pixels with counts 5$\sigma$ above the noise with respect to the average value of their adjacent pixels. With this, we guarantee that the masked pixels are not statistical fluctuations of the Poisson noise but true anomalous fluxes coming from residuals in the masking process.

Finally, we calibrated the reference sky background level. To do this we performed a masking of the diffuse light present in the images. First we created a sum image with the rebinned \textit{g}, \textit{r} and \textit{i} images. Then, we convolved this image with a Gaussian of a large kernel of $\sigma$ = 30 arcsec. After this, we ran \texttt{SExtractor} with a specific setup to {mask all pixels with diffuse emission. With this procedure we manage to leave unmasked all the regions with a minimum flux within the selected field, therefore susceptible of being referenced as the sky background. These empty areas are common to all the individual \textit{g}, \textit{r}, \textit{i} and \textit{z} bands. With this we guarantee that even if there is a residual flux level common to the entire selected field, the relative photometry between bands, that is, the colors, will be robustly determined as they are referenced to the same common area. To improve the sky background determination, we use the routine \texttt{SKY} in IDL to perform this task. This routine makes use of an algorithm of the type MMM (mean-median-mode) to compute a robust sky reference even when the field is overwhelmingly contaminated by diffuse emission, performing an iterative statistical calculation that converges toward a final robust sky background value. In the next sections, we give additional arguments to show the reliability of the calculated background reference levels.}

\section{Galactic cirri: Optical vs. far-IR counterparts}\label{sec:3}

The data processing described in the previous section allows us to isolate the extended diffuse emission present in the optical data. However, it is necessary to discern whether this diffuse emission is due entirely to Galactic dust, or on the contrary, there is another source of diffuse emission in the optical, including residuals of the data processing. Since Galactic cirri appear clearly at far IR wavelengths, any emission by cirri in the optical should have an infrared counterpart, and equivalently, any optical diffuse emission without IR counterpart could be considered not due to cirri. 

\begin{figure*}
  \centering
   \includegraphics[width=1.0\textwidth]{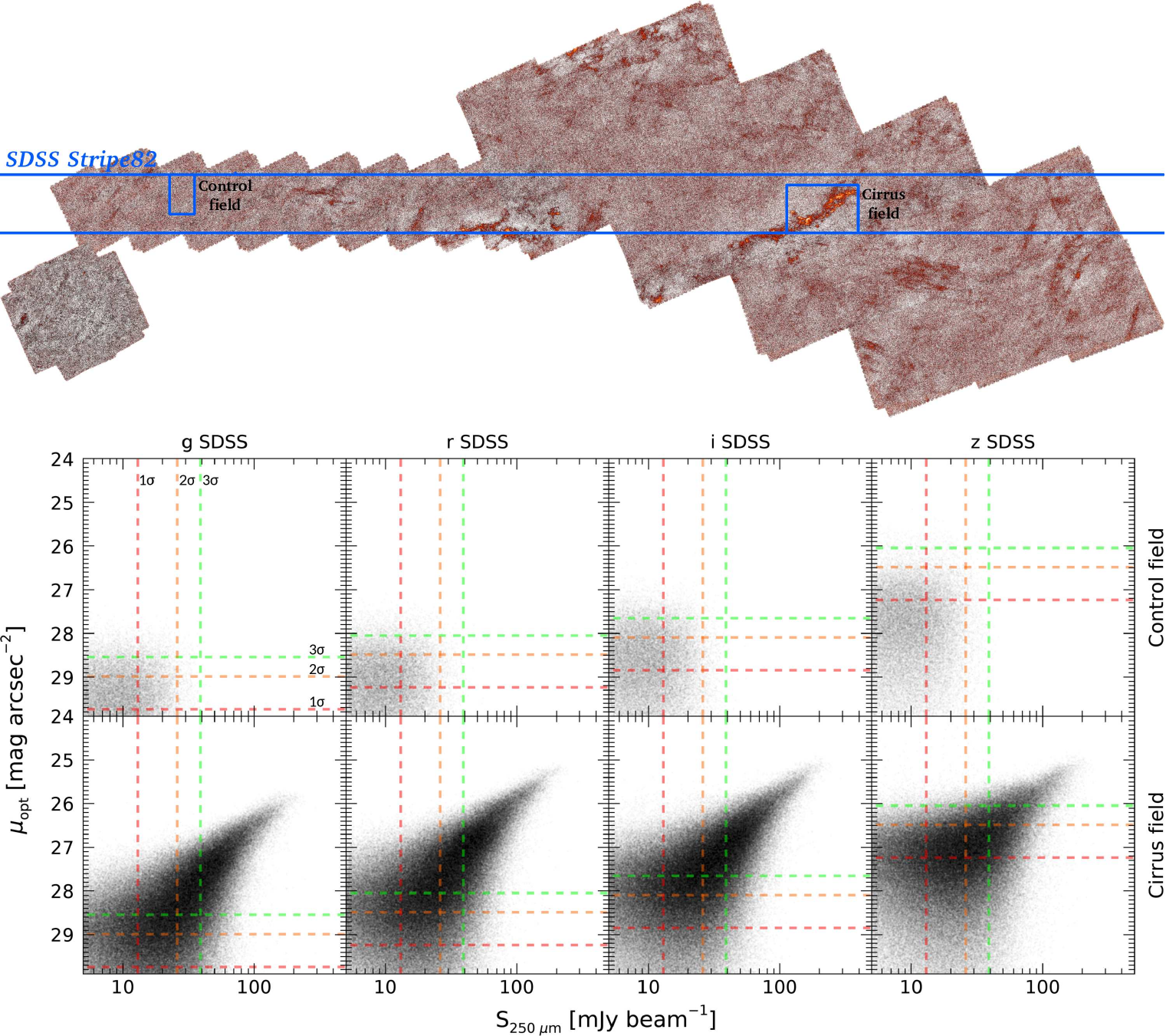}%0.635
    \caption{Optical vs. far IR diffuse emission. In the upper panel is shown the Hershel Stripe82 survey footprint overlapped with the SDSS Stripe82 footprint. The lower panels show the correlation between the diffuse emission in the far IR data and \textit{g}, \textit{r}, \textit{i} and \textit{z} optical bands for the Control field (upper row) and the Cirrus field (lower row). The dashed red, orange and green lines correspond to the 1$\sigma$, 2$\sigma$ and 3$\sigma$ detection limits in each band, respectively. {The effective area of the beam is 0.994$\times$10$^{-8}$ steradians \citep{2013ApJ...772...77V}.}}
   \label{fig:IR_opt}
\end{figure*}

To do that, we performed a direct photometric comparison of the diffuse emission present in the optical data of the IAC Stripe82 Legacy Survey with the diffuse emission found in the far IR in the Herschel Multi-tiered Extragalactic Survey \citep[HerMES;][]{2010MNRAS.409...83L,2013ApJ...772...77V} and the Hershel Stripe82 survey \citep[HerS;][]{2014ApJS..210...22V}. These data cover a wide region matching the Stripe82 area within 350\degree $<$ R.A. $<$ 36\degree \ using the Spectral and Photometric Imaging Receiver (SPIRE) of the \textit{Herschel Space Observatory} at 250, 350, and 500 $\mu$m bands. We used the HerS-HeLMS-XMM-LSS combined SPIRE maps (DR4) publicly available on the Herschel Database in Marseille\footnote{\url{https://hedam.lam.fr/HerMES/}.}. Though this far IR survey is focused on the study of extragalactic sources as infrared-emitting dusty star-forming galaxies or active galactic nuclei, Galactic cirri are detectable in the area mapped by the survey. The relative high spatial resolution of this far IR data compared, for instance, with the IRAS survey, allows a direct photometric comparison of the extended diffuse emission with the optical bands. For the purposes of this work, we used only the 250 $\mu$m band due to its higher angular resolution (pixel scale = 6\arcsec , FWHM = 18.1\arcsec) and optimal wavelength for the detection of the Galactic dust emission.

The top panel in Fig. \ref{fig:IR_opt} shows the HerS-HeLMS-XMM-LSS map at 250 $\mu$m with the footprint of the SDSS Stripe82 survey indicated by horizontal blue lines. While the mapped area has a relatively low presence of dust, some conspicuous filamentary clouds are clearly visible in the 250 $\mu$m image. We selected an area of 3 $\times$ 2 degrees with the highest presence of Galactic cirri, labeled in Fig. \ref{fig:IR_opt} as "Cirrus field". We did not conduct any analysis of additional regions due to the low emissivity of the clouds present in the footprint, both in the far IR and in optical bands, especially in the shallower \textit{z} band. We also selected an area free of Galactic cirri, with a size of $1.0\degree \times 1.5\degree$, as a control field, labelled in Fig. \ref{fig:IR_opt} as "Control field". For the two selected fields, we created five mosaics, one for each of the 250 $\mu$m, \textit{g}, \textit{r}, \textit{i} and \textit{z} bands. The images in different bands were matched astrometrically allowing a pixel to pixel comparison. We performed the data processing described in Sec.~\ref{sec:proccesing} for the optical bands. Additionally, we performed a masking of the 250 $\mu$m image, focused on extragalactic sources leaving unmasked the emission by Galactic cirri. Finally, because the FWHM of the 250 $\mu$m band is 18 arcseconds, we performed a 3-pixel kernel gaussian smoothing to the optical images (pixel scale of 6 arcseconds), ensuring that the optical bands match the resolution of the 250 $\mu$m data.

\begin{figure*}
  \centering
   \includegraphics[width=0.83\textwidth]{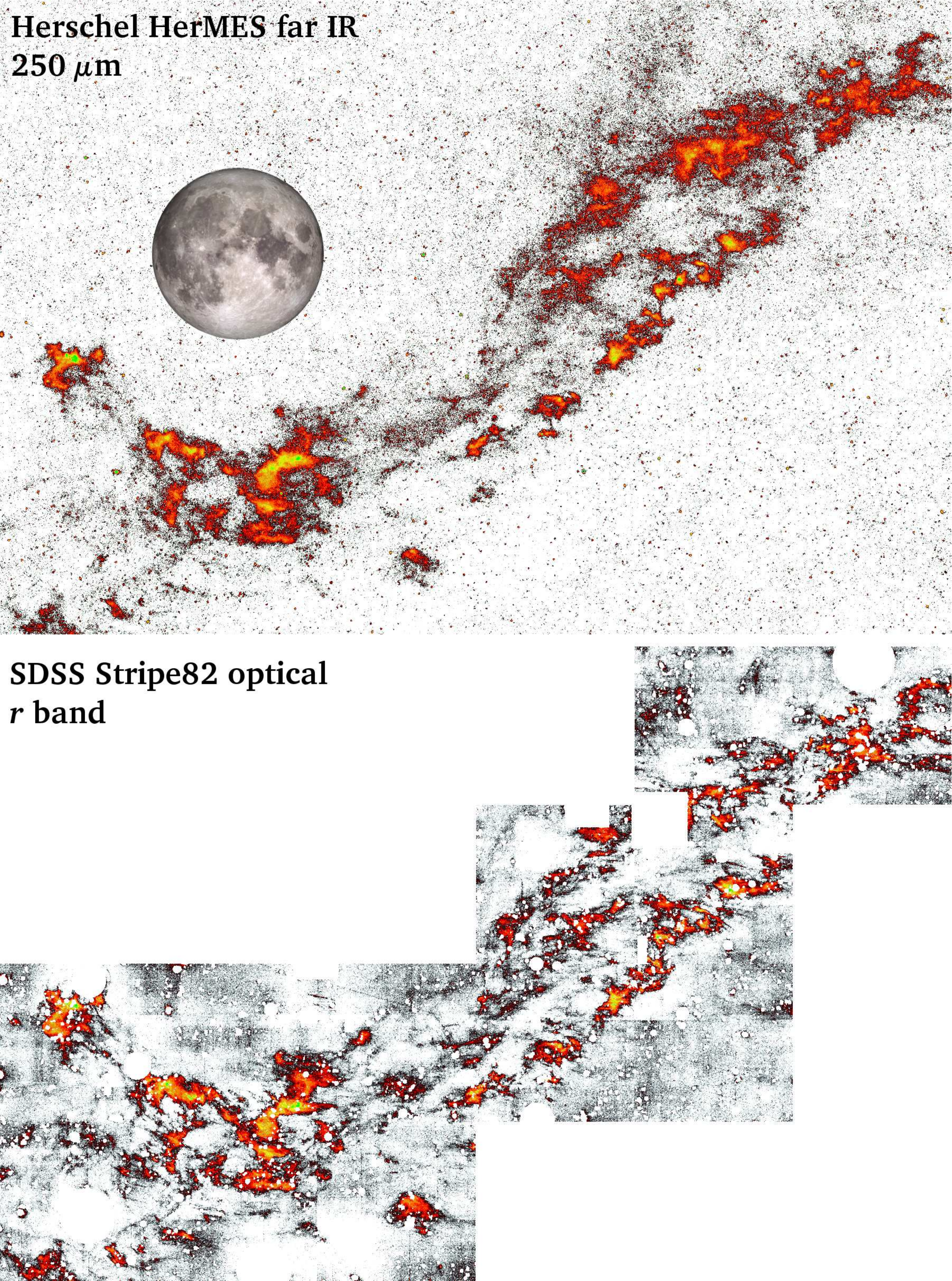}
    \caption{Same Galactic cirrus as observed in the far IR and optical, named as "Cirrus field" (see main text). Top panel: 250 $\mu$m band from the HerMES map. Lower panel: \textit{r} band from the IAC Stripe82 Legacy Survey. The pixel scale for both images is 6 arcsec. An image of the moon (31 arcmin in diameter) is placed as angular scale reference.}

   \label{fig:cirrus_filament}
\end{figure*}

We plot the correlation between the 250 $\mu$m and optical bands in the lower panel of Fig. \ref{fig:IR_opt} for the Control field (upper row) and for the Cirrus field (lower row). Each pixel of the image is a dot in the plot. The detection limits for each band are indicated as dashed lines in red, orange and green, corresponding to the 1$\sigma$, 2$\sigma$ and 3$\sigma$ depth limits, respectively. We used the values of the 250 $\mu$m depth detection limit provided by \cite{2014ApJS..210...22V} corresponding to 1$\sigma$ = 13 mJy beam$^{-1}$. The analysis in the Control Field shows a distribution compatible with Poisson noise in both 250 $\mu$m and optical bands, being the fluctuations of such noise consistent with the detection limits of each band. We find no evidence of optical diffuse emission in the Control field by visual inspection of the images, neither in the correlation plot between the optical and far IR bands, existing only random fluctuations in the sky background due to the drift scan mode of the optical observations, and all of them below the detection limits. Conversely, we find in the Cirrus field a tight correlation between all the optical bands and the far IR emission. While for the \textit{g}, \textit{r} and \textit{i} bands there is a sufficient margin to observe this correlation in a wide range of surface brightness above the detection limits (around 2.5 - 3 mag arcsec$^{-2}$), in the \textit{z} band the magnitude range of detection is significantly smaller due to the shallower depth of this band. Nevertheless, it is still possible to observe such correlation above the 3$\sigma$ limit in areas with surface brightness above 26 mag arcsec$^{-2}$ in the \textit{z} band. It is worth noticing that the depth of the Stripe82 survey is such that the signal to noise of the cirri in the optical images is higher than in the far IR. This can be seen by observing that the cirrus emission is a factor of $\sim$2 further away from the 3$\sigma$ detection limit in the optical bands (except for the \textit{z} band) than for the 250 $\mu$m band. It is also remarkable the similar emission in surface brightness of the \textit{r}, \textit{i} and \textit{z} bands, at least in the brightest regions, being the emission in the \textit{g} band $\sim$ 0.5 mag arcsec$^{-2}$ fainter. This would correspond to \textit{r-i} and \textit{i-z} colors close to 0 mag, while the \textit{g-r} color would be considerably redder. We explore this matter in detail in the next section. 

\begin{figure*}
  \centering
   \includegraphics[width=\textwidth]{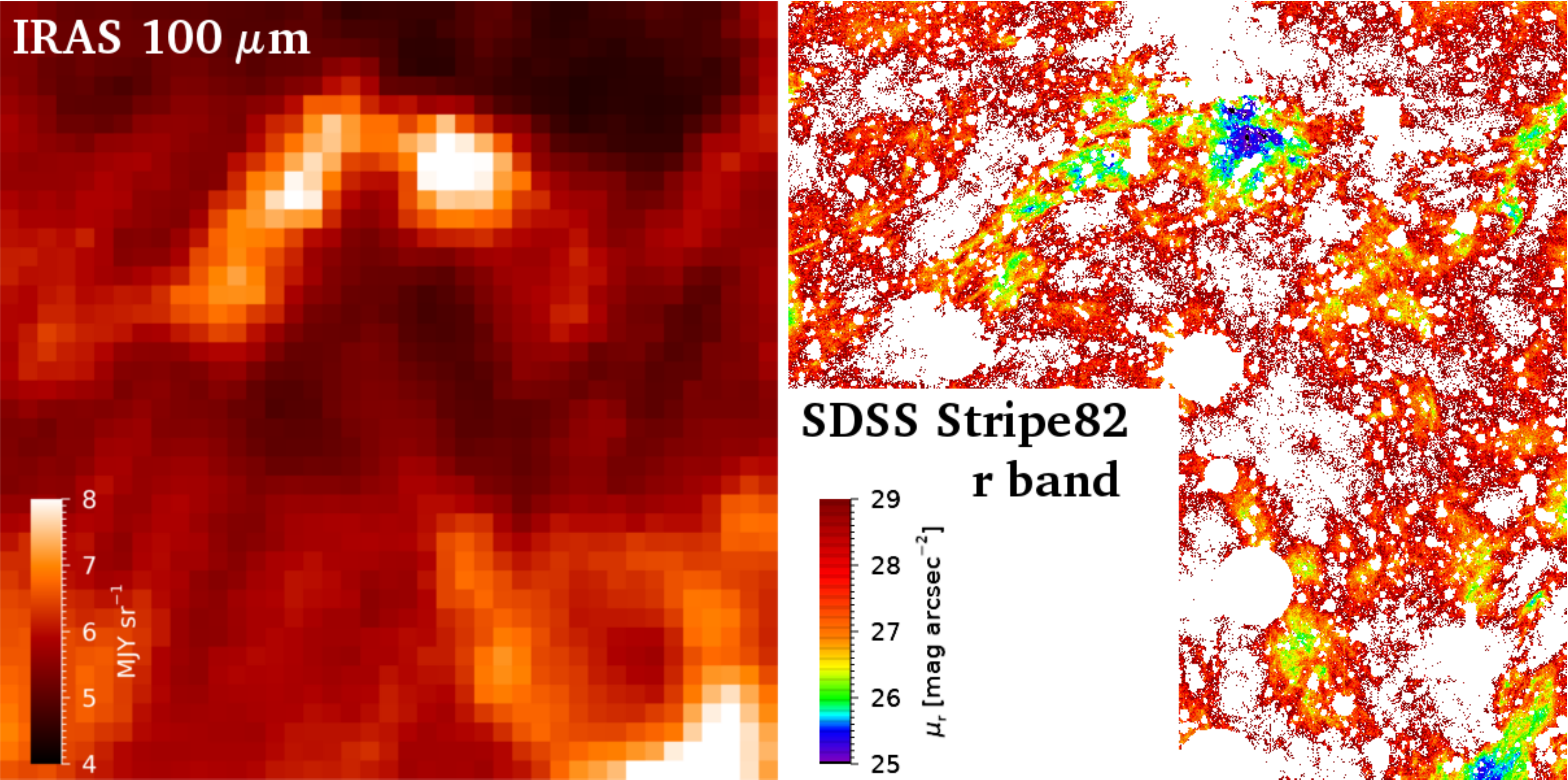}
    \caption{IRAS 100 $\mu$m band image (left panel) and \textit{r} band image (right panel) of the Field 2 presented in Table \ref{tab:fields}. Masked areas appear white in the \textit{r} band. The field has dimensions of 1x1 degrees. The pixel scale is 6 arcseconds in the optical image. A region of this mosaic was used as an example of the image processing in Section \ref{sec:proccesing}, Fig. \ref{fig:Processing1}.}
   \label{fig:cirrus_filamentB}
\end{figure*}

We show in Fig. \ref{fig:cirrus_filament} a visual comparison of the Cirrus field in the far IR 250 $\mu$m band with the processed optical \textit{r} band. The morphology of the diffuse emission in both images is remarkably similar, existing an optical counterpart to what is detected in the far IR. As can be noted, the presence of sky background fluctuations and oversubtraction due to the observational drift scan mode are the main limitations in the recovery of the diffuse emission in the optical dataset. However, these artifacts are below the detection limits. The results in Fig. 6 confirm that the far IR emission by Galactic cirri has a counterpart in the optical. It is worth also emphasizing the extremely low surface brightness of the features detected in this analysis.

The Herschel data allow us to explore the optical counterparts of Galactic dust clouds with a relatively high resolution (FWHM = 18 arcseconds). However, the scarce presence of Galactic cirri in the footprint of the HerS-HeLMS-XMM-LSS maps makes the conclusions obtained so far limited to a particular region, both in Galactic latitude and in IR emission. In order to obtain a more significant sample, we have expanded our analysis to other areas within the Stripe82 region with presence of Galactic cirri using the IRAS 100 $\mu$m full-sky map. Even though the resolution of the IRAS 100 $\mu$m band (FWHM = 5 arcmin) is significantly worse than that of the Herschel SPIRE data, it can still be used to check for the presence of Galactic cirri in the explored areas. We visually compared a set of fields where the data processing in Sec.~\ref{sec:proccesing} has been carried out. The description of the fields and the criteria for selecting them can be found in the next section. We found a good match between the optical data and its far IR counterpart within the resolution limits of IRAS, and more importantly, no obvious optical diffuse emission is detected without a far IR counterpart. We show in Fig. \ref{fig:cirrus_filamentB} an example field of Galactic cirri emission and its counterpart in the 100 $\mu$m IRAS band. This comparison shows the great advantage of deep optical imaging in revealing the morphology of the Galactic dust clouds in a more effective way than the IRAS maps. While the far IR has traditionally been used to detect the presence of interstellar dust, our analysis shows that it is possible to obtain a map of Galactic dust with higher spatial resolution and greater depth using optical data. In summary, the analysis carried out in this section shows that: i) The {instrumental} scattered light field removal process and subsequent masking of the optical images is highly effective in isolating the optical diffuse emission. The IAC Stripe82 Survey ultimate limitation are the sky background fluctuations. Nonetheless, these are below the surface brightness limits of the explored cirri regions; ii) We find no optical diffuse emission without a far IR counterpart, at least down to 28.5 mag arcsec$^{-2}$ (\textit{g} band). Consequently, all the highly extended diffuse emission present in the optical data can be considered due to Galactic dust above that surface brightness; iii) Deep optical data can be used to detect Galactic cirri as well as far IR data. Optical data has the advantage of having a much better spatial resolution.

\section{The optical colors of the Galactic cirri}\label{sec:colors}

The appearance of the Galactic cirri resemble the shape and brightness of extragalactic structures, such as tidal streams or any other stellar diffuse feature in deep optical observations. It is, therefore, imperative to characterize the optical properties of the Galactic dust to identify it against extragalactic sources. In particular, we are interested in exploring any potential difference between the optical colors of the Galactic cirri and extragalactic sources. Assuming that the optical colors of the cirri are due to the light reflected by the surrounding stars, these colors may vary depending on the chemical composition of the dust grains, the dust column density or the average colors of the surrounding stars. In order to explore possible color trends, we selected a number of fields in the Stripe82 region.

The selection of the fields was carried out using only images of the IAC Stripe82 data through visual exploration. With such strategy, we wanted to avoid any potential bias towards selecting diffuse optical emission with obvious IR counterpart. We also tried to select as much as possible isolated cirri, in order to avoid the photometric complexity of having several clouds in the line of sight. In the case of having large areas (several square degrees) susceptible of analysis, we splitted the region into different fields. This was done to provide smaller fields to explore possible color trends. The fields selected under these guidelines are listed in Table \ref{tab:fields} ranging from \# = 1 to 5, and are located in the Stripe82 region between R.A. = [324\degree, 4\degree]. Additionally, we selected areas with strong optical diffuse emission in order to draw possible correlations with the dust column density. These fields are more problematic in terms of the photometric analysis due to the difficulty in the calibration of the sky background level. An extra problem comes from the residuals of the drift-scan observational mode of SDSS. The SDSS reduction pipeline fits a polynomial to the sky emission whose typical length is on the order of several arcmin. Under the presence of any diffuse emission of similar or higher spatial scale, this SDSS algorithm will tend to interpret the diffuse emission as the sky emission itself. This causes a systematic over-subtraction of the data. Nevertheless, the same sky subtraction polynomial fitting is performed in different bands. Hence, for fields with a high contamination and large extension of diffuse emission, we do not consider the absolute value of the surface brightness in each band as reliable. Regardless of this, we show that the derived colors are robust. The higher presence of optical diffuse emission is found in two regions: 326\degree~$<$~R.A.~$<$~330\degree {} (fields ranging from \# = 6 to 10) and 54\degree~$<$~R.A.~$<$~60\degree {} (fields ranging from \# = 11 to 16). The final number of selected fields according with these criteria are 16 (including the "Cirrus Field" analyzed in the previous section as Field \# = 5).

\begin{table*}
\centering
{\small
\caption{Selected fields for the photometric analysis of the Galactic cirri. {Each of the optical color estimations given for each field corresponds to the value obtained by using two different methodologies: the Lorentzian+Gaussian analysis of the signal (upper values) and the flux fraction technique (bottom values). See text for details.}}
\label{tab:fields}
\begin{tabular}{cccccccccccc}

\hline
Field & R.A. (J2000) & Dec. (J2000) & l & b & Area & $<$\textit{g-r}$>$ & $<$\textit{r-i}$>$ & $<$\textit{i-z}$>$ & $<$S$_{100\mu m}$$>$ & $<\tau_{353Ghz}>$ \\

\# & [deg] & [deg]  & [deg]  & [deg] & [deg$^{2}$] & [mag] & [mag] & [mag] & [MJy sr$^{-1}$] & [$\times$10$^{-6}$] \\
\hline
%\multicolumn{10}{c}{Filamentary regions}\\
%\hline
\multirow{2}{*}{1} & \multirow{2}{*}{324.0} & \multirow{2}{*}{0.0} & \multirow{2}{*}{54.7} & \multirow{2}{*}{-35.8} & \multirow{2}{*}{2} & 0.57$\pm$0.04 & 0.12$\pm$0.04 & 0.00$\pm$0.11 & \multirow{2}{*}{3.84} & \multirow{2}{*}{4.18} \\
& & & & & &0.49$\pm$0.05 &  0.18$\pm$0.05  &  0.11$\pm$0.19\\
\hline
\multirow{2}{*}{2} & \multirow{2}{*}{333.0} &  \multirow{2}{*}{-0.75} & \multirow{2}{*}{60.8} & \multirow{2}{*}{-43.6} & \multirow{2}{*}{0.75} & 0.60$\pm$0.04 & 0.05$\pm$0.04 & 0.01$\pm$0.11 & \multirow{2}{*}{5.02} & \multirow{2}{*}{6.06} \\
& & & & & &0.52$\pm$0.04  & 0.13$\pm$0.03  &  0.19$\pm$0.07\\
\hline
\multirow{2}{*}{3} & \multirow{2}{*}{342.75} & \multirow{2}{*}{0.25} & \multirow{2}{*}{71.4} & \multirow{2}{*}{-50.2} & \multirow{2}{*}{1.75} & 0.63$\pm$0.04 & 0.13$\pm$0.04 & 0.10$\pm$0.11 & \multirow{2}{*}{5.59} & \multirow{2}{*}{7.20} \\
& & & & & &0.46$\pm$0.10 &  0.07$\pm$0.10  &  0.12$\pm$0.30\\
\hline
\multirow{2}{*}{4} & \multirow{2}{*}{344.0} & \multirow{2}{*}{-0.25} & \multirow{2}{*}{72.3} & \multirow{2}{*}{-51.4} & \multirow{2}{*}{2} & 0.53$\pm$0.04 & 0.12$\pm$0.04 & 0.08$\pm$0.11 & \multirow{2}{*}{4.54} & \multirow{2}{*}{5.18}\\
& & & & & &0.55$\pm$0.03 &  0.13$\pm$0.02  &  0.11$\pm$0.11\\
\hline
\multirow{2}{*}{5} & \multirow{2}{*}{2.5} & \multirow{2}{*}{-0.25} & \multirow{2}{*}{100.9} & \multirow{2}{*}{-61.3} & \multirow{2}{*}{3} & 0.57$\pm$0.04 & 0.06$\pm$0.04 & 0.04$\pm$0.11 & \multirow{2}{*}{3.97} & \multirow{2}{*}{2.91}\\
& & & & & &0.56$\pm$0.05 &  0.15$\pm$0.04  &  0.02$\pm$0.15 \\
\hline
\multirow{2}{*}{6} & \multirow{2}{*}{326.0} &  \multirow{2}{*}{0.25} & \multirow{2}{*}{56.4} & \multirow{2}{*}{-37.4} & \multirow{2}{*}{2.0} & 0.57$\pm$0.04 & 0.17$\pm$0.04 & 0.11$\pm$0.11 & \multirow{2}{*}{4.04} & \multirow{2}{*}{4.38}\\
& & & & & &0.51$\pm$0.04 &  0.08$\pm$0.05   & 0.07$\pm$0.09\\
\hline
\multirow{2}{*}{7} & \multirow{2}{*}{327.0} &  \multirow{2}{*}{0.0}    & \multirow{2}{*}{56.7} & \multirow{2}{*}{-38.3} & \multirow{2}{*}{2.25} & 0.63$\pm$0.04 & 0.20$\pm$0.04 & 0.02$\pm$0.11 & \multirow{2}{*}{5.82} & \multirow{2}{*}{7.38}\\
& & & & & &0.66$\pm$0.04 &  0.22$\pm$0.03   & 0.02$\pm$0.11\\
\hline
\multirow{2}{*}{8} & \multirow{2}{*}{328.0} &  \multirow{2}{*}{0.0}    & \multirow{2}{*}{57.5} & \multirow{2}{*}{-39.0} & \multirow{2}{*}{2.5} & 0.67$\pm$0.04 & 0.20$\pm$0.04 & 0.07$\pm$0.11 & \multirow{2}{*}{6.19} & \multirow{2}{*}{7.87}\\
& & & & & &0.66$\pm$0.03  & 0.30$\pm$0.02  &  0.13$\pm$0.06\\
\hline
\multirow{2}{*}{9} & \multirow{2}{*}{329.0} & \multirow{2}{*}{-0.25} & \multirow{2}{*}{58.1} & \multirow{2}{*}{-40.0} & \multirow{2}{*}{1.75} & 0.59$\pm$0.04 & 0.17$\pm$0.04 & 0.10$\pm$0.11 & \multirow{2}{*}{5.89} & \multirow{2}{*}{6.78}\\
& & & & & &0.47$\pm$0.07  & 0.26$\pm$0.06  &  0.05$\pm$0.24\\
\hline
\multirow{2}{*}{10} & \multirow{2}{*}{330.0} & \multirow{2}{*}{-0.75} & \multirow{2}{*}{59.1} & \multirow{2}{*}{-40.7} & \multirow{2}{*}{0.75} & 0.57$\pm$0.04 & 0.13$\pm$0.04 & 0.08$\pm$0.11 & \multirow{2}{*}{5.20} & \multirow{2}{*}{4.94} \\
& & & & & &0.57$\pm$0.10 &  -0.01$\pm$0.20 &  0.25$\pm$0.38\\
\hline
\multirow{2}{*}{11} & \multirow{2}{*}{54.95} &  \multirow{2}{*}{0.50} & \multirow{2}{*}{185.6} & \multirow{2}{*}{-41.0} & \multirow{2}{*}{1.35} & 0.61$\pm$0.04 & 0.14$\pm$0.04 & 0.09$\pm$0.11 & \multirow{2}{*}{5.41} & \multirow{2}{*}{6.70}\\
& & & & & &0.63$\pm$0.09  & 0.13$\pm$0.11  &  0.03$\pm$0.15\\
\hline
\multirow{2}{*}{12} & \multirow{2}{*}{55.85} &  \multirow{2}{*}{0.50} & \multirow{2}{*}{186.3} & \multirow{2}{*}{-40.3} & \multirow{2}{*}{1.35} & 0.69$\pm$0.04 & 0.15$\pm$0.04 & -0.10$\pm$0.11 & \multirow{2}{*}{5.98} & \multirow{2}{*}{8.07}\\
& & & & & &0.66$\pm$0.08 &  0.16$\pm$0.04  &  -0.01$\pm$0.11\\
\hline
\multirow{2}{*}{13} & \multirow{2}{*}{56.75} &  \multirow{2}{*}{0.50} & \multirow{2}{*}{187.1} & \multirow{2}{*}{-39.6} & \multirow{2}{*}{1.35} & 0.73$\pm$0.04 & 0.18$\pm$0.04 & -0.05$\pm$0.11 & \multirow{2}{*}{8.28} & \multirow{2}{*}{13.63}\\
& & & & & &0.60$\pm$0.09 &  0.20$\pm$0.03  & -0.04$\pm$0.17\\
\hline
\multirow{2}{*}{14} & \multirow{2}{*}{57.65} &  \multirow{2}{*}{0.50} & \multirow{2}{*}{187.8} & \multirow{2}{*}{-38.9} & \multirow{2}{*}{1.35} & 0.72$\pm$0.04 & 0.17$\pm$0.04 & 0.01$\pm$0.11 & \multirow{2}{*}{9.01} & \multirow{2}{*}{16.15}\\
& & & & & &0.93$\pm$0.05 &  0.12$\pm$0.09  &  0.12$\pm$0.10\\
\hline
\multirow{2}{*}{15} & \multirow{2}{*}{58.55} &  \multirow{2}{*}{0.50} & \multirow{2}{*}{188.4} & \multirow{2}{*}{-38.2} & \multirow{2}{*}{1.35} & 0.83$\pm$0.04 & 0.25$\pm$0.04 & 0.05$\pm$0.11 & \multirow{2}{*}{14.02} & \multirow{2}{*}{25.55}\\
& & & & & &0.71$\pm$0.06 &  0.24$\pm$0.03  &  0.05$\pm$0.11\\
\hline
\multirow{2}{*}{16} & \multirow{2}{*}{59.5} &  \multirow{2}{*}{-0.75} & \multirow{2}{*}{190.4} & \multirow{2}{*}{-38.1} & \multirow{2}{*}{1.0} & 0.92$\pm$0.04 & 0.30$\pm$0.04 & 0.01$\pm$0.11 & \multirow{2}{*}{15.73} & \multirow{2}{*}{28.90}\\
& & & & & &0.84$\pm$0.09 &  0.23$\pm$0.06  &  0.02$\pm$0.17\\
\hline

\hline

\end{tabular}
}
\end{table*}

All the selected fields were processed following the methodology described in Section \ref{sec:proccesing}, including the removal of the stars, masking, rebinning to 6 arcsec pixel scale and calibration of the sky background level. We compared the optical emission of the selected fields with the emission in the 100 $\mu$m IRAS band, and we always found an equivalent counterpart within the spatial resolution limits of the IRAS data. As a tracer of the dust column density we used the average emission values of the 100 $\mu$m IRAS band, listed for each field in Table \ref{tab:fields}.
{To estimate the true average emission of the 100 $\mu$m IRAS band, we have applied the optical data mask, as derived in Section \ref{sec:proccesing}, to the IRAS data in order to eliminate the contribution from any extragalactic source (typically star-forming galaxies). }

\subsection{Methodology for obtaining the optical colors of the cirri}
\label{sec:Methodology}

Performing photometry of extended and extremely faint sources is not a straight forward task. First, the use of specialized software to produce segmentation maps in which to perform automatic photometry of extended astronomical sources is not optimal in this case. Based on the techniques developed previously (that mask all the external sources), we consider that the remaining flux in the images is due to cirri, which will be considered as a single source. However, to obtain the colors of the cirri, the integration of the total flux in the images is not possible due to the presence of oversubtraction and sky background fluctuation regions, which would strongly alter the photometry. In addition, the low signal due to the extremely low surface brightness of the cirri should also be taken into account. Having in mind all the previous considerations, {we have developed a photometric method based on the color distribution of the pixels of the images, in order to obtain an accurate characterization of the colors of the cirri.}

In the case of total absence of sources, that is, purely noise, the flux distribution in each band is described by a Gaussian function due to the Poisson nature of the noise in the images, whose width defines the depth. In building a photometric color, we are performing the flux ratio between both filters and so, the color distribution is described by a Lorentzian function. However, as calculating the magnitude (i.e., applying the logarithm) of the flux values of the pixels, we are taking only the positive values of the distribution. On taking the positive values, the color distribution of the noise in the image is characterized by a Lorentzian distribution centered at $x_{L} = -2.5\times\log(\sigma_{f1})+2.5\times\log(\sigma_{f2})$, where $\sigma_{f1}$ and $\sigma_{f2}$ are the 1$\sigma$ width of the distribution in counts for both bands respectively. In other words, the center of the color distribution (in the AB system) is defined by the difference in depth between both bands.

% Matame camion v2

\begin{figure*}
  \centering
   \includegraphics[width=1.0\textwidth]{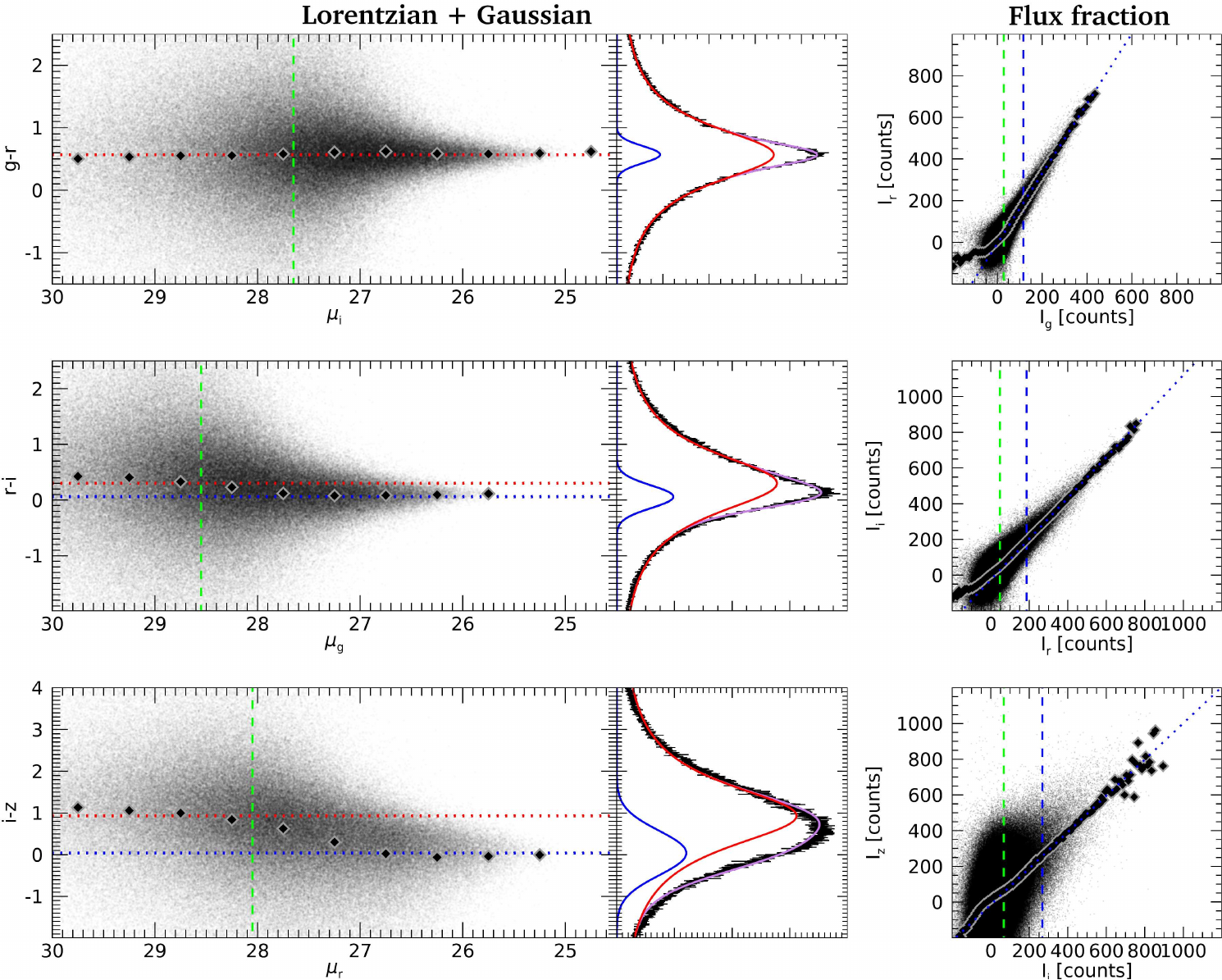}
    \caption{{Visual representation of the methodology for obtaining the colors of the Cirrus field. Left panels: Color vs. surface brightness distribution of the pixels (6 arcsec width). The black diamonds represent the average color in surface brightness bins of 0.5 mag arcsec$^{-2}$. The green dashed line marks the 3$\sigma$ detection limit of the band in the x-axis. The blue and red dotted lines mark the center of the Gaussian and Lorentzian fitted functions respectively ($x_{g}$ and $x_{L}$). The central panels correspond to the color distribution histograms. The black line shows the observed color distribution. The purple line shows the Gaussian plus Lorentzian fitted function, while the blue and red lines show the Gaussian and Lorentzian functions independently. On the right panels, we illustrate the flux fraction method. The correlation between the fluxes in each optical band is shown. The green dashed line marks the 3$\sigma$ detection limit of the band in the x-axis. The blue dashed line marks the 3$\sigma$ detection limit + 1.5 mag arcsec$^{-2}$ of the band in the x-axis (see text for details). The black diamonds represent the average flux of the band in the y-axis in bins of 10 counts for the band in the x-axis. The dotted line indicates the best linear fit.}}

   \label{fig:Diffuse_fit}
\end{figure*}

In our case, the color distribution in the images is a combination of noise and the intrinsic color of the cirri. The color distribution of the cirri is, in principle, unknown. A good approximation is to assume that it is described by a Gaussian distribution. This Gaussianity would appear as the result of intrinsic color variations of the dust combined with random noise. Following this approach, we fit the color distribution of the pixels within an image with a Lorentzian plus a Gaussian function:

$$f(x) \equiv  I_{L} \frac{\frac{1}{2}\Gamma}{(x-x_{L})^{2}+(\frac{1}{2}\Gamma)^{2}} + I_{g}\exp\left(-\frac{(x-x_{g})^{2}}{2\sigma^{2}}\right), $$

where x is the given optical color and \{$I_{L}$,$\Gamma$,$x_{L}$\}, \{$I_{g}$,$\sigma$,$x_{g}$\} are the intensity, width and center of the Lorentzian and Gaussian functions respectively. Thus, the resulting $x_{g}$ will give the average color value of the dust in the analyzed field. As discussed above, the Gaussian width ($\sigma$) will account for intrinsic color variations of the dust. For that reason, the color obtained by this technique must be considered as an average color in the analyzed area. We tested this approach through simulations carried out in the Control Field (no presence of dust). We injected structures of known colors and similar surface brightness to that of the typical cirri analyzed in this work. These simulations show the accuracy in the recovery of the cirri colors using this method. We estimated the following photometric errors: $\Delta$\textit{(g-r)} = 0.04 mag, $\Delta$\textit{(r-i)} = 0.04 mag and $\Delta$\textit{(i-z)} = 0.11 mag. We describe these simulations and further details of the method in Appendix \ref{sec:simulation}. 

In Fig. \ref{fig:Diffuse_fit} we show our analysis applied to the Cirrus field \# 5 in Table \ref{tab:fields}. {In the central panels}, we plot the color histograms of pixels for \textit{g-r}, \textit{r-i} and \textit{i-z}. Overplotted to the histograms are the fits to the distributions using a combination of a Lorentzian plus a Gaussian functions (purple color). The Lorentzian (red) and Gaussian (blue) separated distributions are also plotted to show their relative contribution. As can be observed, the color distributions are very well described with the combination of a Lorentzian plus a Gaussian distributions. We found that in the case of the \textit{i-z} color, there is a slight deviation with respect to the fitting function due to the shallower \textit{z} band and its higher sky fluctuations. Alternatively, {in the left plots}, we show the color vs. surface brightness distributions of the pixels. To avoid artificial correlations, the color distributions are plotted against the surface brightness of a band that is not included in the color. We represent with diamonds the average color values in surface brightness bins of 0.5 mag arcsec$^{-2}$. As expected, the color of brighter pixels is dominated by the typical color of the Cirrus under study (dashed blue line). As we move to fainter regions or pixels, we find a transition to a color dominated by the color of the noise (dashed red line). In the case of the \textit{g-r} color distribution, both the typical color of the cirrus and the color of the noise are very similar and the transition between both regimes can not be well appreciated. Using the methodology proposed in this work, the dust colors in this particular field are \textit{g-r} = 0.57 $\pm$ 0.04 mag, \textit{r-i} = 0.06 $\pm$ 0.04 mag and \textit{i-z} = 0.04 $\pm$ 0.11 mag. It should be noted that these colors agree with the two-dimensional color maps of this cirrus field shown in Appendix \ref{sec:AppendixD} and the trends observed in Fig \ref{fig:IR_opt}.

%{\subsection{Flux fraction method for obtaining the color of the Galactic cirri}}

{In addition to the method proposed in this work, we calculate the optical colors of the cirri by estimating the flux fraction between bands. This method has been widely used in the photometric description of cirri in previous works and it is a standard on color determinations \citep[e.g.,][]{1989ApJ...346..773G, 2010ApJ...723.1549S, 2014ApJS..213...32M}. This method consists on calculating the slope of the correlation in flux between 2 bands. Given 2 bands $A$ and $B$, a linear fitting of the type I$_A$=I$_B\times m +n$ is performed, so the slope $m$ gives the color $A-B$ after its transformation to units of magnitude. This method has the advantage that the color determination obtained is independent of the reference sky background values between bands, which is accounted by the $n$ parameter. However, a drawback of this method is that the color contribution of the noise is not taken into account, so it is necessary to limit the color determination above a certain flux level, leading to a photometric bias of greater or lesser impact. Additionally, the nature of the method assumes that the color is constant, and independent on the flux value in a given band, thus obviating possible correlations between color and flux intensity. We expand on these issues in Appendix \ref{sec:AppendixD}. The right panels of Fig. \ref{fig:Diffuse_fit} show the procedure used to calculate the colors through the flux fraction method. We plot the correlation between the flux values of the bands corresponding to the 3 colors \textit{g-r}, \textit{r-i} and \textit{i-z} for all the pixels. To mitigate the detrimental effect of a flux cut on the correlation, we calculate the average flux value of the band on the y axis on bins for the band on the x axis, plotted as diamonds. We mark the 3$\sigma$ detection limit in the band on the x-axis (the deepest band) as a green dashed line. As can be seeing in Fig. \ref{fig:Diffuse_fit}, the convergence to average dust color is not reached at flux values (or surface brightness) immediately above than the 3$\sigma$ detection limit per band due to the effect of noise from individual bands. For this reason, we define the threshold corresponding to the value of 3$\sigma$ plus the flux corresponding to an added threshold of 1.5 mag arcsec$^{-2}$, which is marked as a blue dashed line in each of the three color correlations. This value of 3$\sigma$ + 1.5 mag arcsec$^{-2}$ is observationally derived from the analysis presented above, in which the convergence to the dust average color is obtained 1.5 mag arcsec$^{-2}$ beyond the 3$\sigma$ detection limit. We, therefore, calculate the correlation of the mean values (diamond symbols) using as errors the bin value for the band in the x-axis and the standard deviation in that bin for the y-axis. We fit a linear least-squares approximation (in one-dimension), giving the slope of the relation, the color, and its error.}

It is worth commenting the strengths of each of the two methods we use in this work. On the one hand, the method of fitting the Lorentzian plus Gaussian functions allows the complete distribution of color to be integrated without making any flux cut. This means that it can be considered free of photometric bias, in contrast to the flux fraction method. Another advantage of the Lorentzian + Gaussian method is that the width of the Gaussian takes into account the expected intrinsic differences in color. Conversely, the flux fraction method is forced to consider a constant color, so any intrinsic color variation will penalize the errors obtained in the mean color of the analyzed field. On the other hand, the flux fraction method is resistant to possible deviations in the sky background calibration, which is undoubtedly the most problematic uncertainty factor of the Lorentzian + Gaussian method. That is why the simultaneous application of these two methods is complementary, since the strength of each of them are the weakness of the other. {In fact, the application of the flux fraction method provides an additional robustness analysis with respect to the estimation of the reference sky background described in Section 2.3. Specifically, when calculating the parameters of the flux correlation between bands of the form I$_A$=I$_B\times m +n$, the parameter n provides interesting information regarding the deviations of the sky flux reference between bands. In the case that the sky background (or flux background) levels are equivalent between bands, the parameter n should be close to 0. This is in fact what we found. We obtain values of n close to 0 within the errors. It is important to note that a value of n~=~0 does not imply that the sky background has been accurately calculated in individual bands, but rather that the sky background values have been calculated in similar regions, or equivalently, between bands. For this reason, the absolute surface brightness values cannot be estimated in an absolute way, but the colors can be obtained, as discussed in Section 2.3. This fact provides additional reliability to the estimation of the sky background described in Section 2.3 and, in general, to the Lorentzian + Gaussian photometric method.}

Finally, we performed the described methodology to all the fields. The total area analyzed is 26.5 square degrees, which corresponds roughly to 10\% of the whole IAC Stripe82 Legacy Survey area. In Table \ref{tab:fields}, we summarize the properties of these fields together with the obtained optical colors for each of the two different methods. The color values of the cirri using the two methods discussed above are generally broadly compatible with each other, suggesting the general reliability of both and the robustness of the colors obtained. Given the compatibility of the results of both methods, we consider the colors obtained with the Lorentzian + Gaussian method as the reference value because of the lower photometric errors in general and the apparently better correlations obtained, as we show in the next section.

\subsection{Optical color trends of the Galactic cirri}

\begin{figure*}
  \centering
   \includegraphics[width=0.84\textwidth]{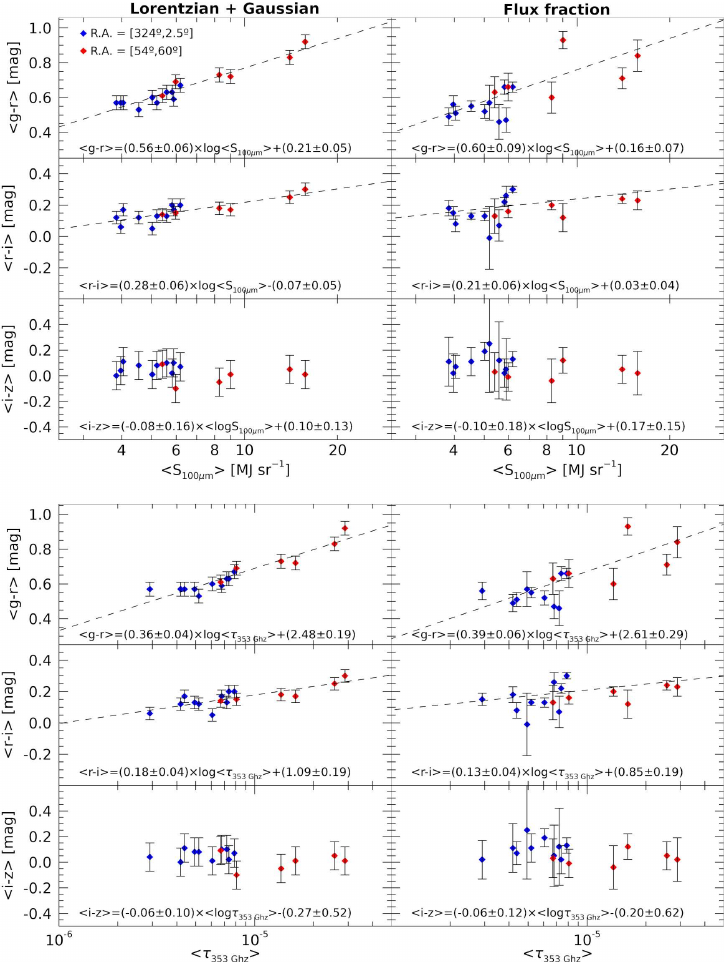}
    \caption{Correlation between the average optical colors of the cirri and the average far IR emission in the 100 $\mu$m IRAS band {(upper panels) and the Planck dust optical depth at 353 Ghz (lower panels)}. The blue and red diamonds represent fields within different right ascension regions as described in the upper left corner. The error bars show the photometric uncertainty for the optical colors. The dashed lines are the best fits as indicated in the bottom equation of each panel. {Panels show the correlations for the two photometric methods used in this work.}}

   \label{fig:Corr_color}
\end{figure*}

\begin{figure*}
  \centering
   \includegraphics[width=1.0\textwidth]{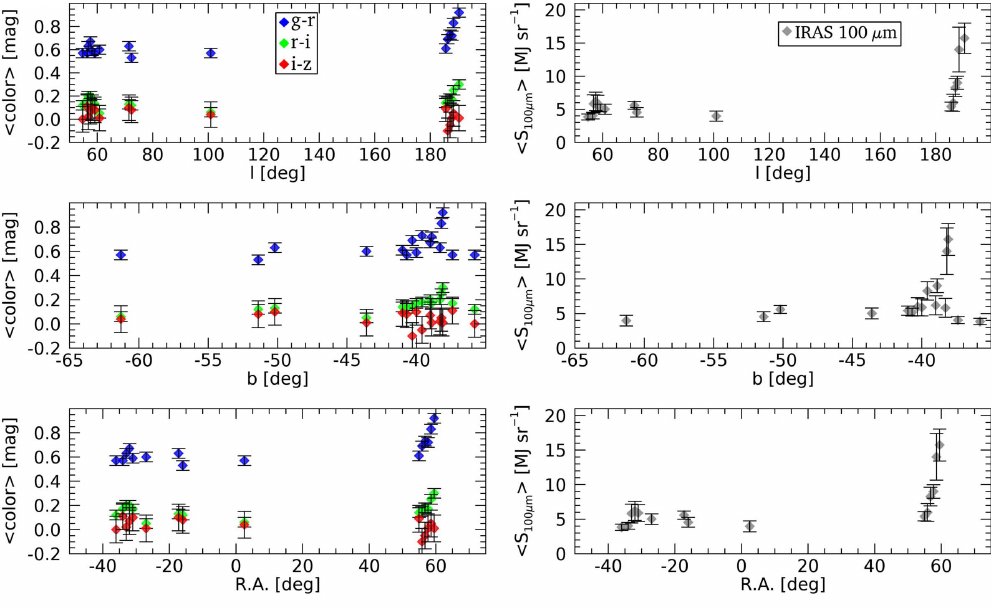}
    \caption{Left panel: Optical colors of the cirri as a function of the location in the sky. Right panel: Same as in the left panel but showing the emission in the IRAS 100 $\mu$m band as comparison.}

   \label{fig:Coords_color}
\end{figure*}

In this section, we explore the optical color trends of the Galactic Cirri. {For this we calculate the correlations of the optical colors of the selected fields with the mean values of 100 $\mu$m emission and optical depth in the selected fields. For the 100 $\mu$m emission we use the IRAS intensity maps discussed in the Section 3. We also use the Planck dust optical depth at 353 Ghz ($\tau_{353 Ghz}$) provided by the Planck collaboration in the Planck Legacy Archive\footnote{\url{http://pla.esac.esa.int/pla/#maps}.}. The average values of the 100 $\mu$m emission and optical depth are calculated in the maps taking into account the masked regions of the optical fields trying to obtain a more reliable comparison. In Table \ref{tab:fields}, we list these values. We assume the error for the average values of 100 $\mu$m and $\tau_{353 Ghz}$ as negligible, due to the large number of pixels to average that value in the selected fields: $\sigma/\sqrt{N}~\approx$~0}. {For the optical colors we use the tabulated values and errors provided in Table \ref{tab:fields}.} 

In Fig. \ref{fig:Corr_color}, we plot the {these correlations} for the selected fields { for each of the two methods used in the calculation of the colors of the cirri}. In both cases, we find a clear correlation between the far IR emission {and optical depth of the dust}, and the optical colors. This correlation is steeper for bluer optical colors, being the one of the $g-r$ color the steepest. We do not find any obvious trend between the far IR emission and the \textit{i-z} color within the error bars. {The slopes of the linear correlations have compatible values between both methods.} {We find very similar trends in the case of $\tau_{353 Ghz}$, indicating a clear correlation between the optical colors of cirri and the dust column density.} To explore any potential trend of the dust colors with the position in the sky, we separate the different fields by coordinates. Fields with right ascension ranging between 324 $\deg$ to 2.5 $\deg$ (fields \# 1 to 10) are plotted in blue while fields with a right ascension ranging between 54 $\deg$ to 60 $\deg$ (fields \# 11 to 16) are plotted in red. We do not find any obvious trend in this correlation as a function of the right ascension coordinates. 

{An interesting exercise would be to compare our measured colors in highly contaminated regions by dust with the spectrum of the diffuse Galactic light provided by \cite{2012ApJ...744..129B}. This spectrum was calculated by averaging SDSS sky spectra in areas with S$_{100 \mu m}$ < 10 MJy sr$^{-1}$. When convolving this spectrum with the SDSS filter response, we obtain the following colors: \textit{g-r} = 0.84 mag and \textit{r-i} = 0.34 mag. These values are not compatible with those found in our work, being much redder. The color properties of the Galactic dust found in our work can be also compared with those found by \citet{2013ApJ...767...80I} also using broad-band filters. We have in common with this work the \textit{g} band. They, however, have \textit{R} band instead of \textit{r} band. Nonetheless, we can compare our \textit{g-r} color with their \textit{g-R} color. The range of study in \citet{2013ApJ...767...80I} is between 1 to 7 MJy sr$^{-1}$ at 100 $\mu m$. We can take 4 MJy sr$^{-1}$ as an average value to make the comparison with our colors. At this IR intensity, they get g-R~$\sim$~0.49 mag, which is very comparable to our $g-r$~$\sim$~0.55 mag. Unfortunately, their range of variation in terms of the IR emission is very modest to explore a potential trend of the optical color as the one we find in Fig. \ref{fig:Corr_color}. However, we can attempt such comparison using the values presented by \cite{1989ApJ...346..773G} using the B-R color at different 100 $\mu m$ intensities. These authors show this color for the ranges 3-7 MJy sr$^{-1}$ and 7-18 MJy sr$^{-1}$. We assume a representative average value of 5 and 13 MJy sr$^{-1}$ to represent both intervals. \cite{1989ApJ...346..773G} find a $\Delta$(B-R)~$\sim$~0.5 mag between both IR intensities, while we find  $\Delta$(g-r)~$\sim$~0.23 mag. This is qualitative consistent, taking into account that the slope of the optical color versus the intensity at 100 $\mu m$ seems to increase at bluer colors}.

As an additional analysis, we plot in Fig. \ref{fig:Coords_color} the optical colors of the cirri vs. their celestial coordinates in right ascension and galactic latitude and longitude. We did not conduct this analysis for the declination because the Stripe82 region subtends a narrow declination range of -2.5 $\deg$ to 2.5 $\deg$. For comparison, we also plot the average 100 $\mu m$ emission. We do not find any trend between the cirri colors and the celestial coordinates within the explored regions, being the color variations of the cirri linked to the emission in the 100 $\mu$m IRAS band.

\subsection{Dust vs. extragalactic features colors}

\begin{figure*}
  \centering
   \includegraphics[width=1.0\textwidth]{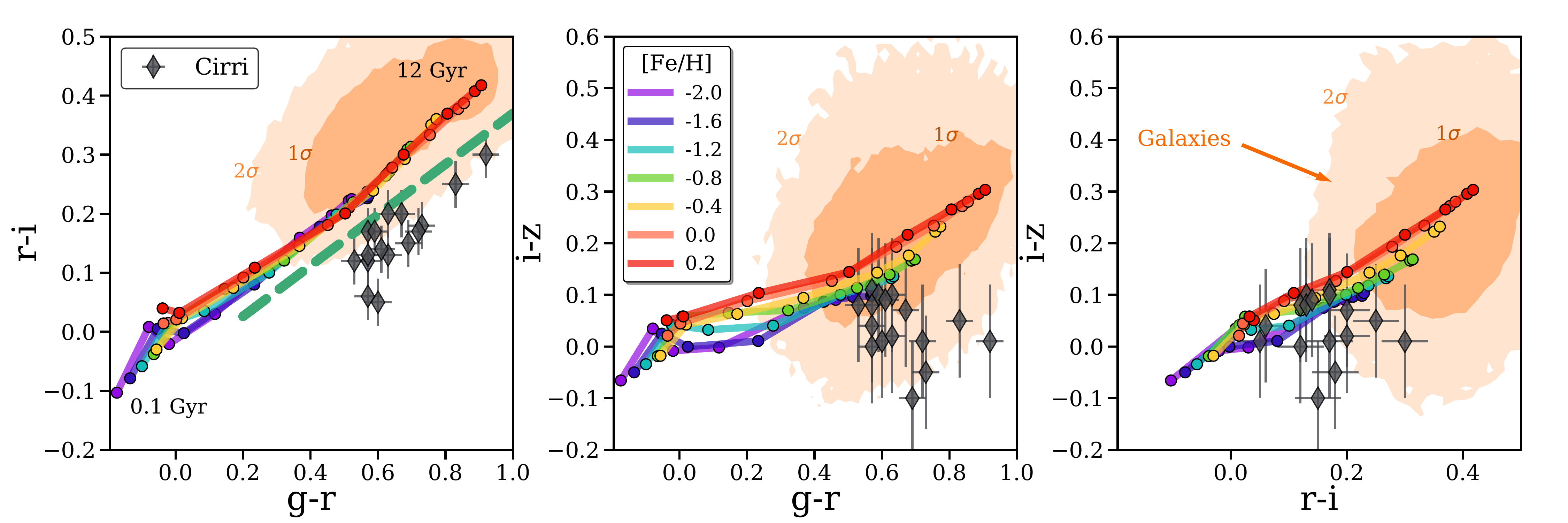}
    \caption{Color-color diagrams of the Galactic cirri and the extragalactic sources. Black diamonds correspond to the cirri colors with their corresponding error bars. Overplotted are E-MILES single stellar population models color-coded for different metallicities at each of the following ages: 0.1, 0.2, 0.5, 1, 2, 5, 10 and 12 Gyr. We also plot the color values of the pixels of real galaxies (see text). The light orange region encloses 95\% (2$\sigma$) of the pixels, while the dark orange encloses 68\% (1$\sigma$). {The green dashed line in the \textit{g-r} vs \textit{r-i} map indicates the separation between the region compatible with dust (below the line) and extragalactic sources (above the line).}}

   \label{fig:Dust_colors}
\end{figure*}

The goal of this work is to explore whether there is any potential difference between the optical colors of the Galactic dust and extragalactic sources. In Fig.~\ref{fig:Dust_colors}, we plot the measured colors of the cirri in the selected fields together with the colors from the E-MILES single stellar population models \citep[][]{2016MNRAS.463.3409V} for different ages and metallicities. We also include the colors of real extragalactic sources in the images. For this, we used the dust-free Control field discussed in Section 3 and we selected pixels belonging to galaxies (extended sources with stellarity parameter less than 0.2) in the original images at the original pixel scale of 0.396 arcseconds. The color of the galaxies is shown as the contour regions in the color-color maps, {which is useful as a means of displaying the color dispersion of individual pixel values from real sources}. We plotted the 1$\sigma$ and 2$\sigma$ density contours. We note that the average colors of real extragalactic sources differ slightly from that of single stellar population models due to the non evaluated k-correction of sources at different redshifts. Additionally, the relatively high dispersion found is consequence of: 1) the presence of multiple stellar populations in real galaxies, 2) the photometry using the original pixel scale of 0.396 arcsec (the cirri colors were obtained averaging its emission at larger regions) and 3) the redshift scatter. It is worth mentioning here that the shallower depth of the z-band translates into a greater uncertainty of the i-z color. {The optical colors of low surface brightness galaxies (susceptible to being easily confused by dust in fields with high dust contamination) as the ones explored by \cite{2017MNRAS.468.4039R} are also within the region enclosed by the orange contours in Fig. \ref{fig:Dust_colors}}. 

While for the \textit{r-i} vs \textit{i-z} map the colors of the cirri overlap with the colors of the single stellar population models and real observed galaxies, for the \textit{g-r} vs \textit{r-i} and \textit{g-r} vs \textit{i-z} diagrams, the colors of the cirri are well differentiated from the single stellar population models and the real extragalactic sources. {We rule out making an analytical estimate of which degree of overlap between the colors of the extragalactic sources and the colors of the cirrus clouds on the \textit{g-r} vs \textit{r-i} map, since there is no single way to define an analytic color distributions for extragalactic sources, being dependent on its distance, color, signal to noise, etc.}

{\subsection{Detection of Galactic cirri using optical colors: a test on the HSC-SSP survey}}

{The good separation in the  \textit{g-r} vs \textit{r-i} diagrams of the colors of Galactic cirri against the colors of extragalactic sources offers a potential tool to break the confusion between them based exclusively on the use of optical bands. Therefore, it is interesting to carry out a test, using independent data, to explore whether we can discern between dust and extragalactic sources based exclusively on optical data. }

{To do this, we have chosen the Hyper Suprime-Cam Subaru Strategic Program data \citep[HSC-SSP;][]{2018PASJ...70S...4A}. This is done for several reasons. First, the HSC-SSP is  deeper by around 1 mag arcsec$^{-2}$ than the IAC Stripe82 Legacy Survey. Second, it has  all the required photometric bands (\textit{g, r, i, z}) used in this work. Additionally, the most recent public data release \citep[PDR2;][]{2019PASJ...71..114A} has significant improvements regarding the stability of the sky background and oversubtraction around sources, so it allows a reliable study of diffuse emission such as Galactic dust. }

\begin{figure*}
  \centering
   \includegraphics[width=0.75\textwidth]{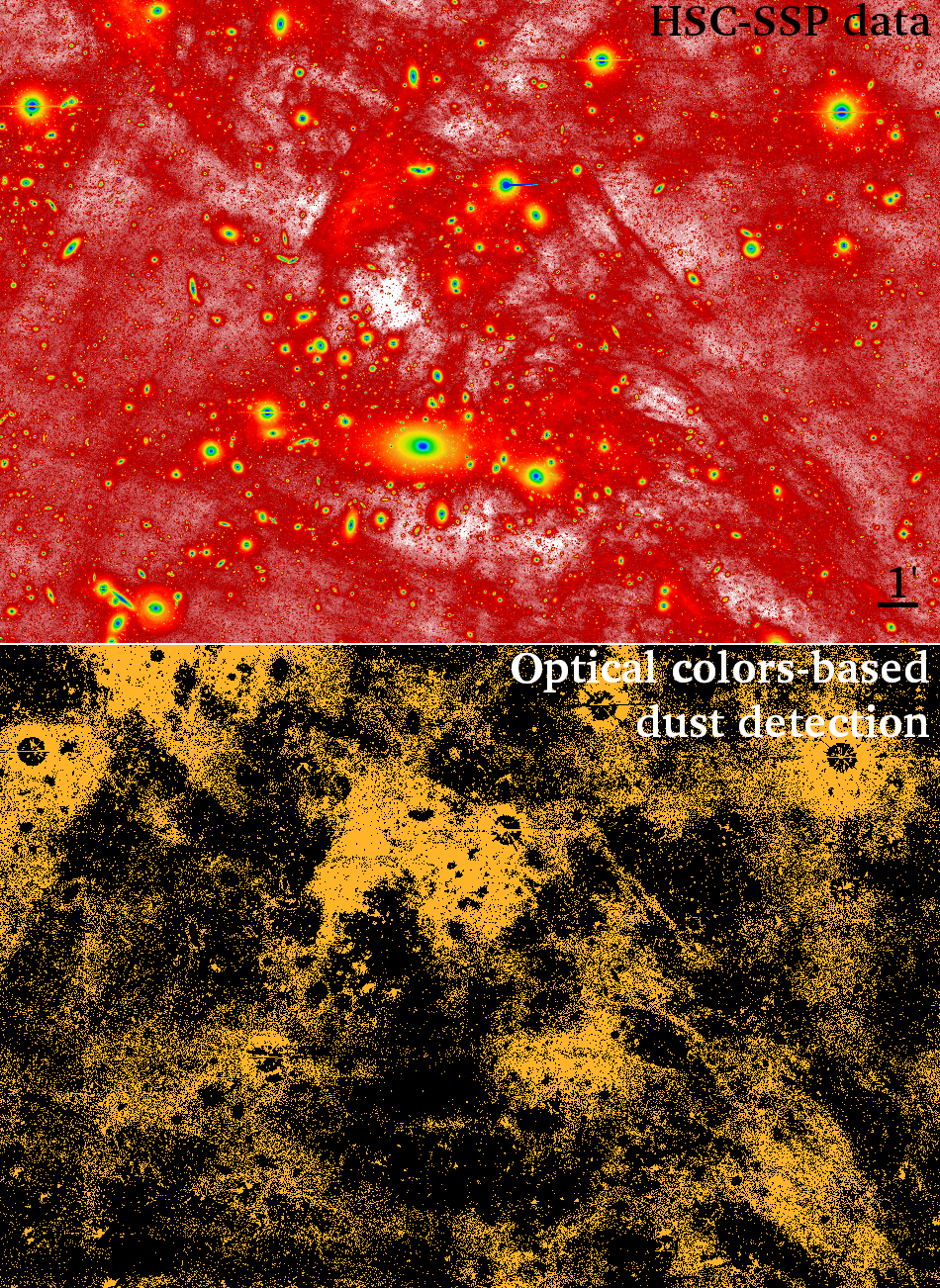}
    \caption{{Central region  of the Abell 2457 galaxy cluster (26.5$\times$18.2 arcmin) analyzed using our dust identification pipeline (see text for details). Upper panel: Sum of the \textit{g, r} and \textit{i} bands with a color scale tuned to obtain an optimal contrast between sources of different surface brightness. Bottom panel: Same region as the upper panel showing those pixels compatible with dust (yellow pixels) according with our analysis.}}
   \label{fig:Dust_detection}
\end{figure*}

\begin{figure*}
  \centering
   \includegraphics[width=0.75\textwidth]{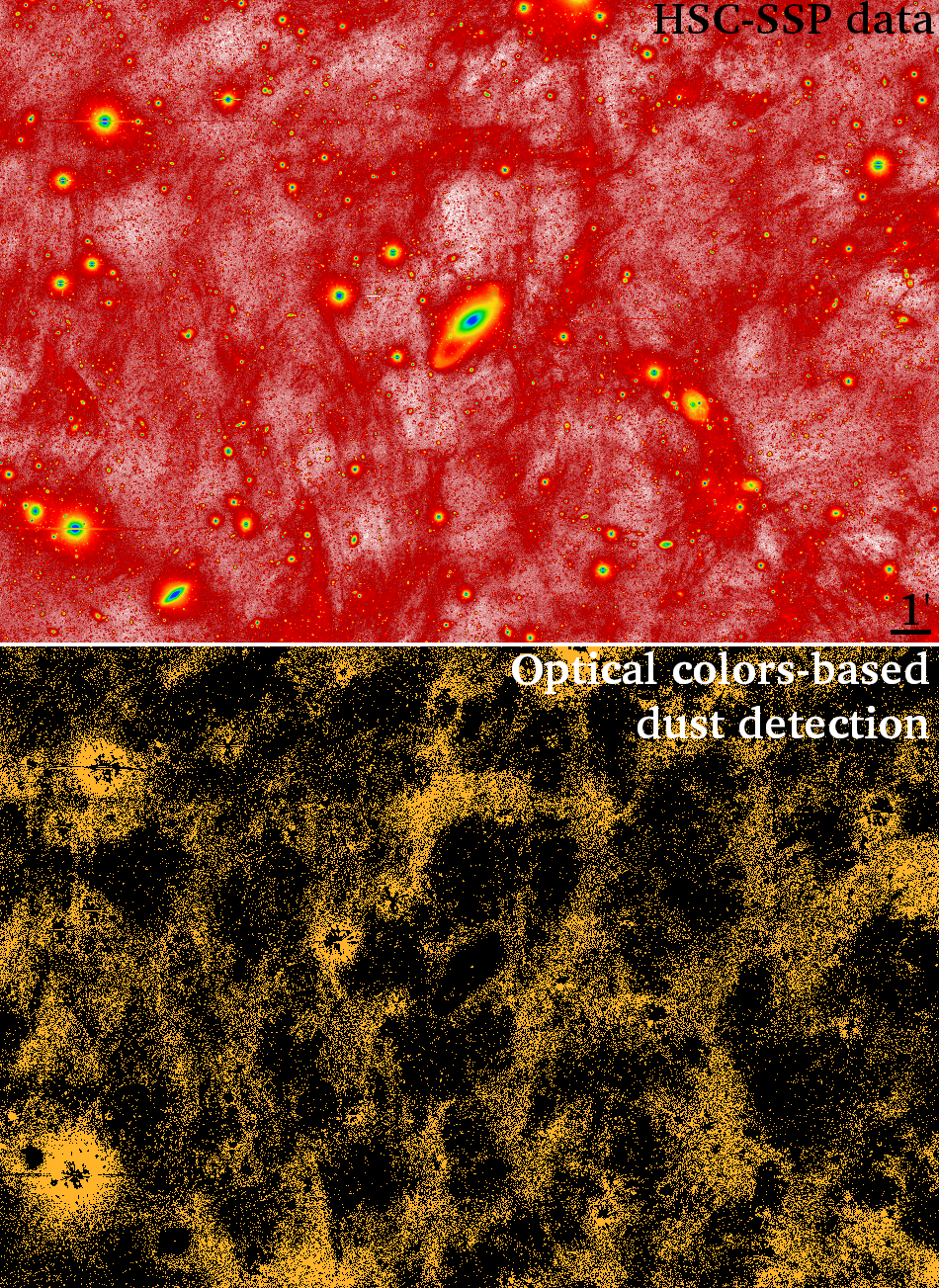}
    \caption{{Region of 26.5$\times$18.2 arcmin centered over the CGCG379-015 galaxy analyzed using our dust identification pipeline (see text for details). Upper panel: Sum of the \textit{g, r} and \textit{i} bands with a color scale tuned to obtain an optimal contrast between sources of different surface brightness. Bottom panel: Same region as the upper panel showing those pixels compatible with dust (yellow pixels) according with our analysis.}}
   \label{fig:Dust_detection2}
\end{figure*}

\begin{figure*}
  \centering
   \includegraphics[width=0.75\textwidth]{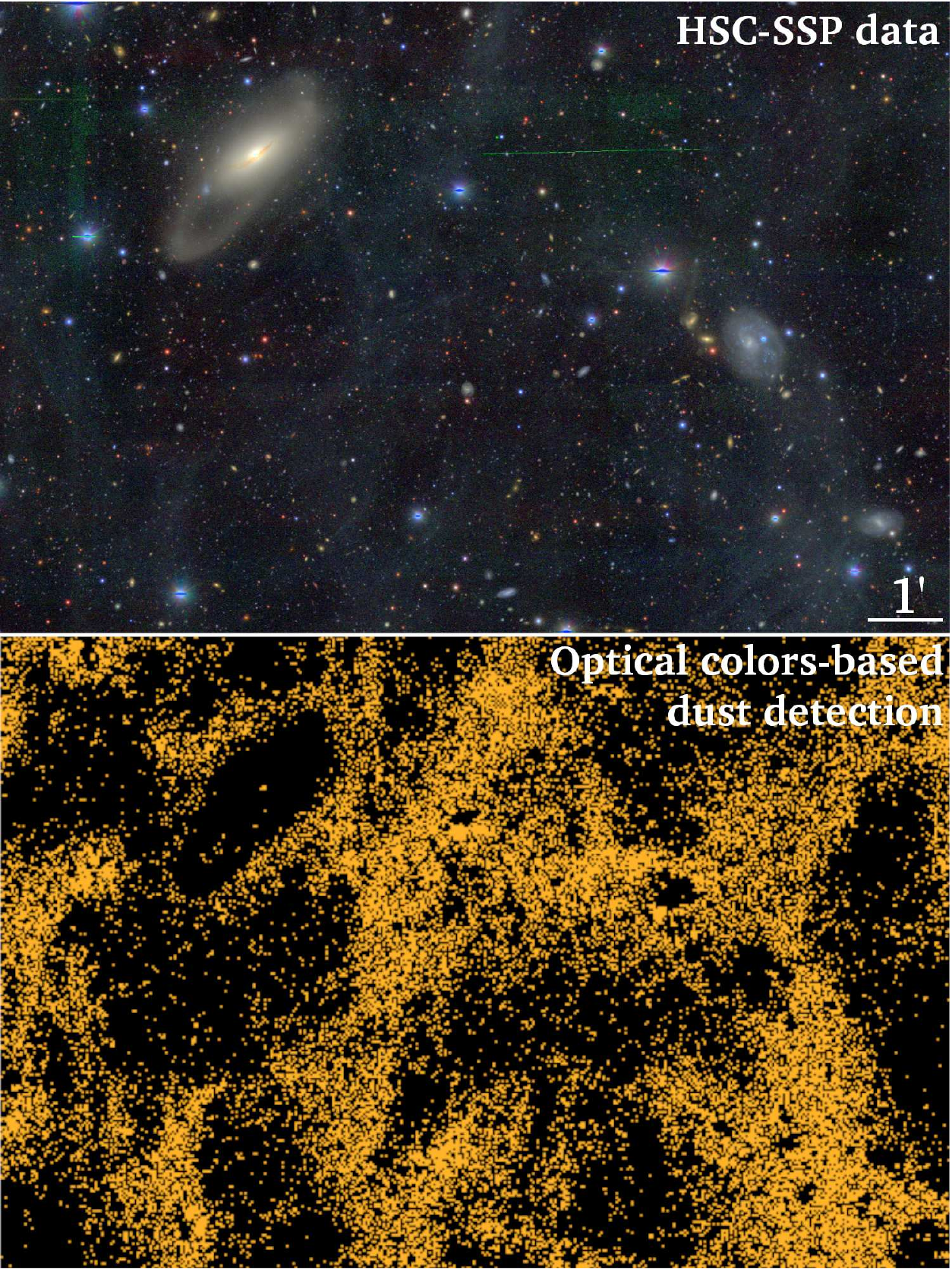}
    \caption{{A zoom-in (12.5$\times$8.5 arcmin) to the region shown in the previous figure showing the CGCG379-015 galaxy and some low surface brightness galaxies. Upper panel: Color composed image using the \textit{g, r} and \textit{i}. Bottom panel:  Same region as the upper panel showing those pixels compatible with dust (yellow pixels) according with our analysis.}}
   \label{fig:Dust_detection3}
\end{figure*}

{Given a set of filters, we estimate the colors of each individual pixels. As according to our color-color plots, the best color combination to segregate between dust and extragalactic emission is given by \textit{g-r} vs \textit{r-i}. Therefore, we use the information provided by that diagram. A pixel is identified as compatible with dust emission if it satisfies:}

\begin{equation}
     (r-i) < 0.43\times (g-r) -0.06.
    \label{eq:dust}
\end{equation}

{For this test, we do not perform any particular processing on the data as masking or modeling and subtraction of the stars but just a rebinning of the images obtaining the subsequent increase of the signal-to-noise in order to have a better reliability and avoiding being dominated by the color of the Poisson noise.}

{We have selected two representative fields to perform this test. These regions have been selected on the basis of containing both extragalactic sources and dust regions, with the addition that they are clearly visually distinguishable from each other. With this, we avoid having to resort to complementary data in the IR. It is worth noting that IRAS data would not be useful here due to its poor spatial resolution, especially compared to the HSC-SSP resolution (0.168 arcsec/pix, average FWHM = 0.6 arcsec in the $i$ band). Moreover, there is no Herschel data of these selected fields. In the upper panel of Fig. \ref{fig:Dust_detection}, we show the first of the fields. This region contains a large number of galaxies of different sizes, morphologies and colors, along with a large amount of diffuse emission, both extragalactic such as the intracluster light of this galaxy cluster (Abell 2457), as well as an apparent high presence of Galactic dust in the line of sight. In the bottom panel of Fig. \ref{fig:Dust_detection}, we show the detection map derived from our criteria in Eq. \ref{eq:dust}. As it can be seen in the detection image, the color criteria does indeed separate the dust filamentary structures from extragalactic sources, tracing remarkably well the morphology of these diffuse regions. Additionally, regions that are clearly galaxies in the field do not appear in the detection image. Something similar can be observed in the second field shown in Fig. \ref{fig:Dust_detection2}. This region was selected for having many of the features that are commonly mistaken with dust in deep data. In the central region we can find the galaxy CGCG379-015 that clearly exhibits a tidal stream in the deep HSC-SSP data. We can also find some low surface brightness galaxies. When applying our dust separation criteria, we are able to identify dusty regions while all extragalactic sources are not selected, showing how well the color differentiation works. It is worth noting the great visual similarity of the colors of the low surface brightness galaxies to the colors of the cirrus clouds present, as can be seen in Fig. \ref{fig:Dust_detection3}. Despite this, our criterion works efficiently appearing these galaxies as "holes" in the dust detection maps. It is worth noting that the {instrumental} scattered light from some bright stars appears as false positives on dust detection maps. A more rigorous analysis should model and subtract all the stars in the field, as we did to obtain the photometric characterization of the cirrus clouds in Section \ref{sec:proccesing}. It is also worth commenting that for this test we have only used the \textit{g-r} vs \textit{r-i} colors, ignoring the \textit{z} band due to its low signal to noise. Despite using only 3 bands, the results effectively show which regions are dominated by dust emission using exclusively optical data.}

\section{Conclusions and discussion}

We have carried out a comprehensive study of the Galactic cirri within the Stripe82 region using data from the IAC Stripe82 Legacy Survey \citep{2016MNRAS.456.1359F}. Through a detailed characterization of the PSF and using custom made techniques particularly designed for studying deep optical data, we have demonstrated that it is possible to characterize the optical colors of the interstellar dust. Both the subtraction of the {instrumental} scattered light from the stars and the exhaustive masking of all the sources are essential for a correct analysis of the dust clouds. We analyzed a total area of 26.5 square degrees. We found a spatial correspondence between the optical emission and the emission of the dust cold component as traced by the far IR bands in Herschel and IRAS bands. Additionally, we did not find any source of optical diffuse emission without an IR counterpart.

The techniques developed in this work allow us to accurately measure the average colors of the Galactic cirri in areas of few square degrees. We found a strong correlation between the average optical colors and the 100 $\mu$m band average emission, {similarly observed in optical depth maps ($\tau_{353 Ghz}$) of the Planck Space Observatory}. This correlation is steeper for the \textit{g}-\textit{r} color than for the \textit{r}-\textit{i} color. We did not find such a correlation for the \textit{i}-\textit{z} color within the photometric uncertainties, much higher for the shallower \textit{z} band. The correlation between the cirri optical colors and the dust column density is a remarkable result that has not been identified previously \citep[see a summary of previous works by][Table 3]{2013ApJ...767...80I}. The previous lack of detection of these correlations could be related with the presence of the {instrumental} scattered light by the stars. In this work, this source of contamination has been carefully subtracted. {Additionally, the high-quality photometry of the Galactic cirri provided in this work could serve as robust observational data with which to constrain models of the interstellar medium \citep{2001ApJ...548..296W,2004ASPC..309...77G}. In particular the data presented in this work may provide better predictions for the properties of the dust albedo as they are poorly constrained by current observations. We further provide a field that might be of interest for constrains of the albedo properties in Appendix \ref{sec:nebula}.}

It is worth exploring which physical mechanisms are able to produce such a correlation between the optical color and the far IR emission for the Galactic dust, as shown in Fig. \ref{fig:Corr_color}. A naive scenario is that the reflected optical light from a given dust cloud could be suffering extra extinction (and so, extra reddening) if there is an additional dust component along the line of sight. Therefore in areas with a higher dust column density, the colors would become redder. Additionally, the blue colors of the cirri found by \cite{2017ApJ...834...16M} in the Virgo Cluster (where the dust contamination by cirri is moderate) and the redder colors found by \cite{2016ApJ...826...59W} in a conspicuous and dense cloud near the M64 galaxy (where the dust contamination by cirri is severe) are in general terms compatible with this scenario. We note that these works subtracted the {instrumental} scattered light by the stars, as we did in this work. {In any case, we do not attempt an explanation of the physical processes responsible for the optical colors and its correlation with the IR. We hope that future works in this regard will find the observational data provided here useful.}

{The optical colors of the dust appear well differentiated from the colors of extragalactic sources with a characteristic lower \textit{r-i} value for a given \textit{g-r} in comparison with extragalactic sources. Taking advantage of this fact, we have probed whether we can identify the dust features in images of the HSC-SSP. The results presented in this work are very promising, showing that the colors of the dust found in the IAC Stripe82 Legacy Survey, can be applied to a different survey, independent of the specific data set considered.}

Finally, it is also worth briefly discussing the potential of different optical surveys in the study of Galactic dust. In this work, we have used data from the IAC Stripe82 Legacy Survey to study wide regions with presence of dust. The imminent arrival of the LSST could be extremely useful in the study of the photometric properties of the Galactic dust. The presence of six photometric bands (\textit{u, g, r, i, z} and \textit{y}), the great depth of the final coadded data and the large explored area (30,000 square degrees of sky) are excellent characteristics for the detection and characterization of the Galactic dust. {To this we can add the IR bands that Euclid will provide}. However, without an excellent reduction of the data preserving low surface brightness sources, and an exquisite characterization of the PSF at large radius to model and subtract the {instrumental} scattered field by stars, the study of extended diffuse emission will be compromised. Assuming that the final images meet the quality required to study the optical diffuse emission, it could be possible to conduct a systematic study of the Galactic dust using this superb optical dataset. The main advantage with respect to surveys in the far IR will be the much better spatial resolution and depth. This would allow to trace the dust and create maps with unprecedented quality. We note that the use of other wavelengths for the detection of dust with a higher spatial resolution have already been proposed, for example by \cite{2015A&A...579A..29B} using data in the UV, which could be complementary to the optical data of the LSST.

\begin{acknowledgements}
{We thank the referees by their detailed revision of the manuscript that helped to improve its quality.} The authors of this paper also acknowledge support from grant AYA2016-77237-C3-1-P from the Spanish Ministry of Economy and Competitiveness (MINECO) and from IAC project P/300624, financed by the Ministry of Science, Innovation and Universities, through the State Budget and by the Canary Islands Department of Economy, Knowledge and Employment, through the Regional Budget of the Autonomous Community. We acknowledge support from the Fundación BBVA under its 2017 program of assistance to scientific research groups, for the project "Using machine-learning techniques to drag galaxies from the noise in deep imaging". JR acknowledge financial support from the grants AYA2015-65973-C3-1-R and RTI2018-096228-B-C31 (MINECO/FEDER, UE), as well as from the State Agency for Research of the Spanish MCIU through the "Center of Excellence Severo Ochoa" award to the Instituto de Astrofísica de Andaluc\'ia (SEV-2017-0709). I.T. acknowledges financial support from the European Union’s Horizon 2020 research and innovation programme under Marie Sklodowska-Curie grant agreement No 721463 to the SUNDIAL ITN network. We thank Timothy Brandt for sharing the spectrum of the diffuse galactic light with us. We thank Adolf Witt for interesting suggestions about our work. We thank Jorge Sánchez Almeida and David Valls-Gabaud for interesting discussions about the results. We thank Aaron Watkins for complementary analysis about the galactic dust colors and Juergen Fliri for his excellent work on the construction of the IAC Stripe82 Legacy Survey. We thank Lee Kelvin for sharing his code to compute beautiful color composed images. JR thanks Anastasia Khamatullina for interesting discussions. We thank the participants of the workshop "Exploring the Ultra-Low Surface Brightness Universe" hosted in the ISSI headquartes, Bern, Switzerland on January 2017, in which preliminary result were shown. { We also thank the participants of the meeting "The low surface brightness Universe as seen by LSST" hosted in Sesto, Italy on January 2020 for interesting discussions.}
\end{acknowledgements}

%-------------------------------------------------------------------

\begin{appendix} 

\section{Surface brightness limits of the data}\label{sec:Depth}

We describe the process for obtaining the surface brightness limits of the data. In Fig. \ref{fig:histo_counts} is shown the histogram of counts of a masked image of the IAC Stripe 82 Legacy Survey for all the \textit{u, g, r, i} and \textit{z} bands. As can be seen, the distributions follow an almost perfect gaussian shape as all the non masked pixels are dominated by the Poisson noise of the data. This Poisson noise is what limits the depth of the data. Fitting a gaussian function we obtain its standard deviation ($\sigma$). Therefore, the surface brightness limit at the 3$\sigma$ level in pixel to pixel variation can be defined as:

$$\mu_{lim} (3\sigma_{pix\times pix})=-2.5 \times log\left(\frac{3\sigma}{pix^{2}}\right) + Zp, $$

where pix is the pixel size in arcseconds and Zp is the zero point of the data. In other words, this is the detection threshold in surface brightness at 3$\sigma$ for a source with an angular size equivalent to that pixel size. However, the typical size of celestial sources is considerably larger. For this reason, the limiting surface brightness at the original pixel scale is not representative of the real detection threshold for extended sources. In addition, each particular survey has its own pixel scale, so this value is not comparable among different surveys or datasets. To obtain the surface brightness limit at any given angular scale, the above histograms of counts have to be obtained at the required angular scale. It could be done by rebinning the image and obtaining the standard deviation of this new distribution. However, the residuals of the masking process after the rebinning and the substantial reduction in the number of spatial elements make this option problematic, specially in images with high contamination or small area. Consequently, the practical way to obtain the surface brightness limit at a given angular scale is assuming that the flux of the noise follows a normal distribution. Then, the Gaussian standard deviation on a different angular scale will transform as: 

$$\sigma({\Omega})=\sigma \sqrt{\frac{pix^2}{\Omega^2}},$$

where $\Omega$ is the desired pixel scale box of the new definition. With it, we can define the surface brightness limit for any angular scale as: 

$$\mu_{lim} (3\sigma_{\Omega \times \Omega})=-2.5 \times log\left(\frac{3\sigma}{pix\times \Omega}\right) + Zp.$$

Traditionally, an angular scale of 10$\times$10 arcseconds boxes has been used as representative of the typical extended source when exploring nearby galaxies. Using this definition, we obtain for the data used in this work: $\mu (3\sigma;10''\times 10'')$ = 28.0, 29.1, 28.6, 28.2 and 26.6 mag arcsec$^{-2}$ for the \textit{u, g, r, i} and \textit{z} bands respectively. It is worth noting that the limiting depth is strongly dependent on the angular size or criteria to define it. For instance, the previous values in $\mu_{lim} (3\sigma;10''\times10'')$ would transform as $\mu_{lim} (1\sigma;10''\times10'')$ = 29.2, 30.3, 29.8, 29.4 and 27.8 mag arcsec$^{-2}$ or $\mu_{lim} (1\sigma;60''\times60'')$ = 31.1, 32.3, 31.7, 31.3 and 29.8 mag arcsec$^{-2}$ for the \textit{u, g, r, i} and \textit{z} bands respectively.

\begin{figure}
  \centering
   \includegraphics[width=0.47\textwidth]{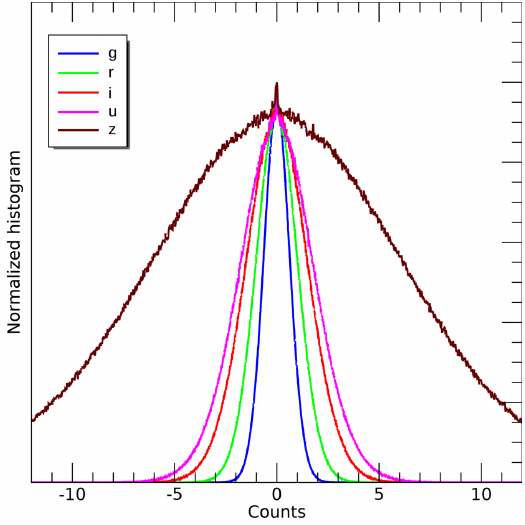}%0.635
    \caption{Distributions of counts for a 0.5$\times$0.5 degrees masked image at the original pixel size of 0.396 arcsec in the five SDSS bands. The histograms in each band follow a normal distribution with a gaussian shape. All bands own a similar zero point (Z$_p$) of 28.3291 mag, as provided by the IAC Stripe 82 survey. We note the shallower \textit{z} band evidenced by a wider distribution.}

   \label{fig:histo_counts}
\end{figure}

\section{PSF characterization}\label{sec:PSF}

The quality of the modeling of the {instrumental} scattered light produced by the stars relies on the accuracy achieved in the characterization of the PSF and the subsequent accuracy at fitting this PSF model to the stars. Bearing this in mind, we have made a characterization of the PSF taking into account possible adverse effects that can penalize a correct PSF model. The commonly used techniques, such as dedicated observations of a very bright star, can introduce systematic effects as it is based in unique observational conditions. This could translate in a different full width at half maximum (FWHM) in the Core of the PSF, or even fluctuations of the PSF wings at large radii \citep[][]{2013JGRD..118.5679D}. Additionally, this kind of "external" characterization of the PSF is not feasible to perform in data from multiple coadded surveys. The reason is that the images of the survey are the coadd of several multi-epoch single exposures, so the PSF of a single exposure could be different of the PSF in the coadded survey. Consequently, a good alternative for obtaining a high quality and realistic PSF is to use the stars in the science images themselves.

In this section, we describe the process for obtaining the PSF model used in this work. This is done by the stacking of many different stars. That "averaged star" is considered the PSF model of the data. To conduct the procedure described here, we use data from the IAC Stripe 82 Legacy Survey \citep{2016MNRAS.456.1359F}, therefore, the PSF models produced here will be those of this survey. Nevertheless, this procedure is general and can be used for any dataset of similar characteristics.

\subsection{PSF Wings}

\begin{figure*}
  \centering
   \includegraphics[width=0.8\textwidth]{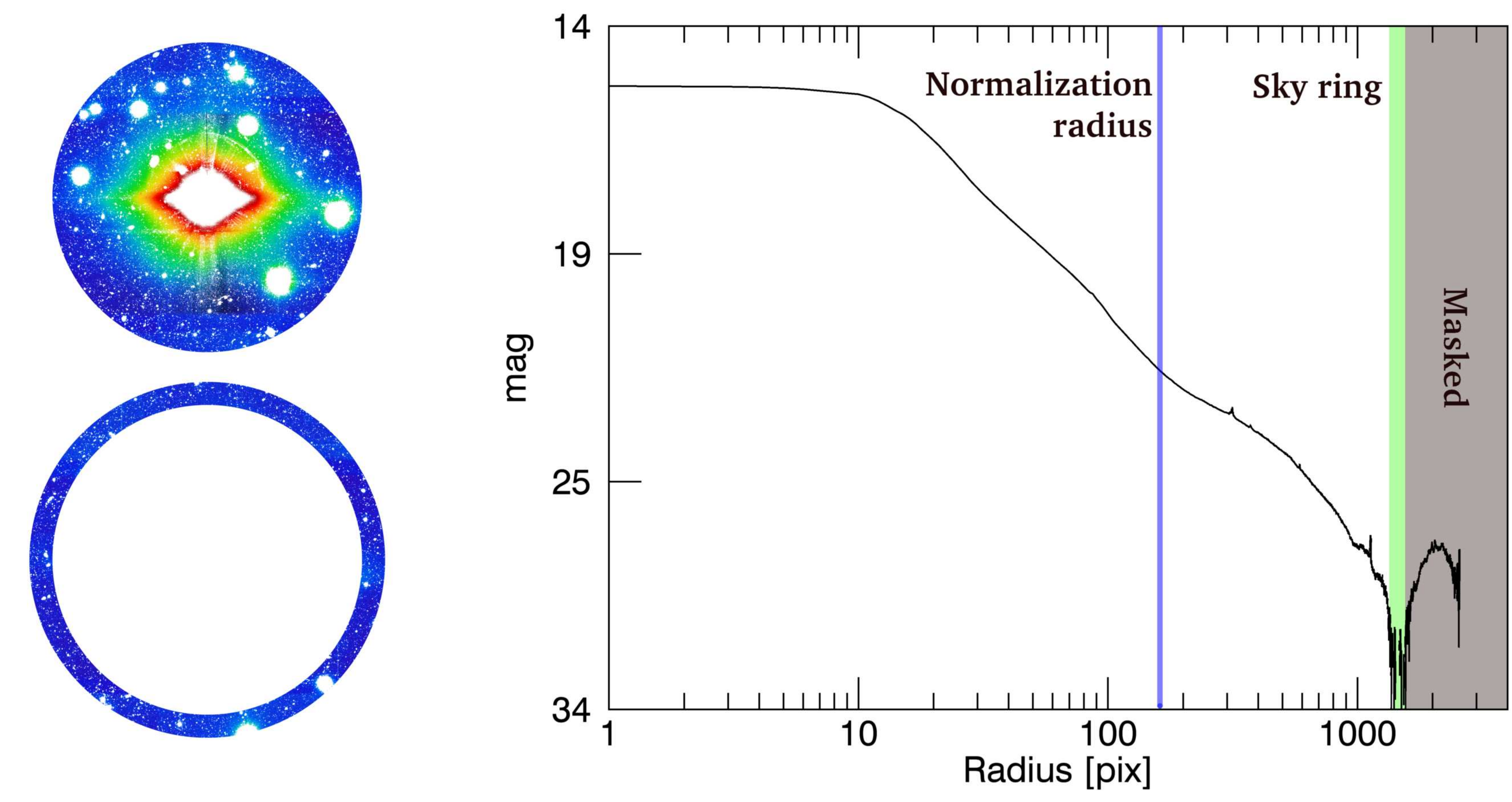}%0.635
    \caption{Schematic representation of the correction process for the construction of the Wing PSFs in the case of oversubtraction around very bright stars. Upper Left: An image of of a very bright star which has its surrounding sky over-subtracted; Bottom left: the sky ring used to calibrate the sky that will be added to the image of the star. Right: Profile of the star showing the different regions.}

   \label{fig:star_annulus}
\end{figure*}

To be able to explore the effect of the {instrumental} scattered light at very low surface brightness, it is necessary to characterize the PSF to the largest possible spatial extent. In the present work, we have built the PSFs up to a radial distance of 720 arcseconds (12 arcmin). Such large PSFs imply that we are able to characterize and remove the {instrumental} scattered light down to extremely faint surface brightness. In order to maximize the reliability of the PSFs, specially at large radius, the selected stars are taken in a region with low contamination by Galactic cirri. The area chosen is the Stripe82 region within the Hers82 survey (avoiding the cirri clouds discussed in Section 3). Due to the precision required, accurate centering of the stars is necessary. The coordinates of bright saturated stars, provided by either SDSS or by creating a catalog running \texttt{SExtractor} \citep[][]{1996A&AS..117..393B}, do not have the desired precision. We use as central coordinates those of the USNO-B catalog of stars \citep[][]{2003AJ....125..984M} and we select stars that have null error in the coordinates according to the USNO-B (J2000 epoch) catalogue. Stars matching this criteria in the selected area are many ($\approx$ 1200 stars with R $<$ 12 mag). However, stars with (R $<$ 5 mag) are only a few. The scarcity of very bright stars forces us to stack a large number of stars in a wide range of luminosities. To combine stars with very different fluxes, we need to scale the different stars to a common flux (i.e., we need to normalize them) so they can share a common profile. The use of aperture photometry for the normalization of the individual stars is ruled out. The stars in this magnitude range are saturated in SDSS, so the flux obtained by using a fixed aperture is not reliable. For this reason we make use of the profile of each star to perform the normalization. 

We proceed as follows. For each star in our preselected catalog, we created an image stamp in each band with an odd number of pixels\footnote{We use an odd number of pixels to create the PSF in order to have a central reference pixel in the 2-D final image} of 720 arcsec using the software \texttt{SWarp} \citep{2002ASPC..281..228B} centered in the coordinates of the star according to the USNO-B catalogue. We masked all the external sources based on the detection map from \texttt{SExtractor} using the \textit{r-deep} band (combination of \textit{g}, \textit{r} and \textit{i} bands). Then we performed a visual inspection to discard unreliable stars. Reasons for discarding a star are the presence of adjacent sources of similar luminosity, saturated rows or columns or any other artifact. Extremely bright stars are especially problematic, in fact, all stars with R $<$ 5 mag were discarded. A common problem in SDSS images is the presence of ghosts in fields adjacent to bright stars. Hence, we performed a manual masking of the ghosts for each star. From our original preselection of 535 stars, we ended up with 399 stars in the magnitude range 5.35 $<$ R $<$ 10.55, with 90\% of the stars having R $>$ 8 mag and 75\% R $>$ 9 mag. Then, we computed the profiles for all the stars. 

Another problem to handle is the oversubtraction of light around bright stars. Although the IAC Stripe82 Legacy Survey is built on the basis of preserving the lower surface brightness structures, a slight over-subtraction is detected around very bright stars. Due to the high precision required for the construction of the PSF, this must be taken into account. To minimize this issue, we developed an automatic routine in which any possible oversubtraction is softened. This is made by detecting changes in the slope of the profiles of the stars that could be indicative of oversubtraction. If oversubtraction is detected, we measured the average sky value in a ring located farthest from the star in the oversubtracted region and we add this value back to the whole stamp. The region of the stamp with radius larger than the selected oversubtracted area is masked (see Fig. \ref{fig:star_annulus} for a visual example). Then, the whole 2-D image is normalized by a multiplicative factor obtained from the flux of the star in the profile between 40 and 60 arcseconds, so each normalized star has the same flux in this region. We choose this range to avoid the central internal reflection of the star, located within the inner 40 arcseconds. We used a maximum radius of 60 arcseconds to make sure that the signal to noise ratio is reliable for the faintest stars. Once all the stars are normalized in flux we proceed with the stacking. We obtained the combined PSF averaging the 2-D images of the stars, computing the mean of each pixel with an iterative rejection based on a gaussian distribution. We rejected pixel values beyond the 2$\sigma$ value. Additionally, for the low signal regions ($>$ 300 arcseconds for the \textit{g, r} and \textit{i} bands and $>$ 100 arcseconds for the \textit{u} and \textit{z} bands) we forced a minimum rejection of the 15\% of the maximum and 15\% of the minimum pixel values. Due to the high luminosity of the stars to construct the Wing PSFs, the inner region appears saturated. Therefore, it is necessary to construct an equivalent Core PSF to characterize this inner central region.

\subsection{PSF core}

The photometry of stars is usually based on PSFs whose size is not larger than few arcseconds. To obtain these small PSFs, the \texttt{PSFEx} package \citep[PSFEx;][]{2011ASPC..442..435B} is commonly used. However, for the construction of the inner region of our PSFs, we used a technique similar to that of the Wing PSFs, this is, the stacking of a large number of stars from the images. The stars were extracted from an area of 2.5\degree$\times$2.5\degree centered at coordinates RA = 25.75, Dec. = 0. This region of the sky has a low contamination by Galactic dust, and importantly has a low number of very bright stars, therefore, low-luminosity stars are weakly influenced by the {instrumental} scattered light from brighter stars. For all the bands, we selected stars in this region using \texttt{SExtractor} with stellarity factor greater than 0.8 and a magnitude range between 15.5 and 16.5 mag. This magnitude range selects the brightest non-saturated stars. The number of selected stars for the stacking depends on the band, being around 600 stars.

We proceed as follows. We created a stamp of 201 pixels ($\approx$ 80 arcsec) using \texttt{SWarp} centered at the coordinates retrieved from \texttt{SExtractor} of each star and we masked all the external sources. As in the case of very bright stars used for the construction of the Wing PSFs, having the most accurate measurement of the central coordinates of the stars is crucial to build a reliable Core PSF. We discarded the use of coordinates from catalogues. Instead, we fit a Moffat function to each star using \texttt{IMFIT} \citep[][]{2015ApJ...799..226E} and we resampled the stamp using the coordinates retrieved by the Moffat best fit. We recalibrated the sky of each star at a radius larger than 80 pixels. To perform the normalization of each star, we measured the flux of the profile from 9 to 11 pixels ($\approx$ 3.5 to 4.5 arcsec). Finally, we obtained the Core PSF by stacking all the stars using a resistant mean with a 2$\sigma$ rejection.

\subsection{Junction of the PSFs}

To obtain the complete PSFs, we performed the junction of the Core with the Wing PSFs for each of the \textit{u, g, r, i} and \textit{z} bands. The main difficulty of this process is the normalization and the relative sky reference of the Core PSF in relation to the Wing PSF. By construction, the sky measured around the stars for creating the Core PSFs (region beyond 80 pixels) is calibrated over a region where there is flux of the Wing PSF, although underneath the noise. This implies having to perform a junction with 2 free parameters, as they should match in flux in the junction point but also they should have a common absolute sky reference, which is unreliable to obtain due to the low signal to noise of the Core PSF at large radius (where the sky reference has been calibrated for the Wing PSF). To solve this problem, we performed the junction of Core and Wings by solving the following set of equations.

$$ Core (R1) = a\times Wing (R1) + b. $$
$$ Core (R2) = a\times Wing (R2) + b. $$

Where R1 and R2 are two different radial distances with enough signal to noise in both the Wing and Core PSFs radial profiles. Hence, solving these equations will give us the normalization value ($a$) and sky level ($b$) to be applied to the Core PSF for a perfect matching previous to the junction. We use different R1 and R2 radial distances for each band trying to avoid the internal reflections and reaching the best signal to noise as possible for the final PSFs \citep[see][for further details]{Infante}. After the normalization of the Core PSFs relative to the Wing PSF we performed the junction. Pixels less than R1 away from the center are referenced to the Core PSF, and pixels with a larger radius are referenced to the Wing PSF, producing the complete PSF. We show the profiles for the complete PSFs of each band in Fig. \ref{fig:PSFs_profiles}.

\begin{figure}
  \centering
   \includegraphics[width=\columnwidth]{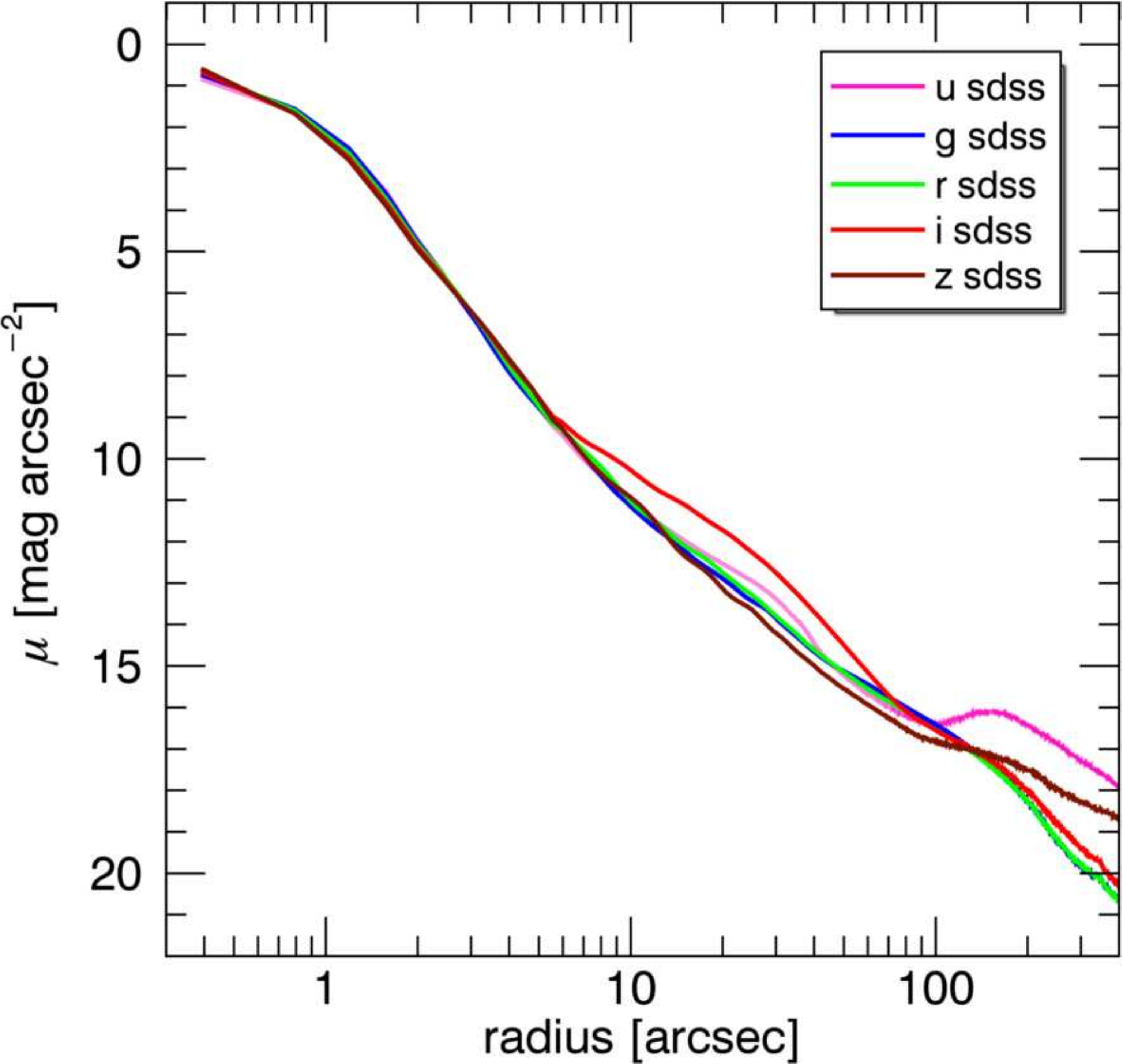}%0.635
    \caption{Radial profiles of the IAC Stripe82 PSFs obtained and used in this work.}

   \label{fig:PSFs_profiles}
\end{figure}

\section{Estimation of photometric errors for the Galactic cirri diffuse emission.}
\label{sec:simulation}

\begin{figure*}
  \centering
   \includegraphics[width=\textwidth]{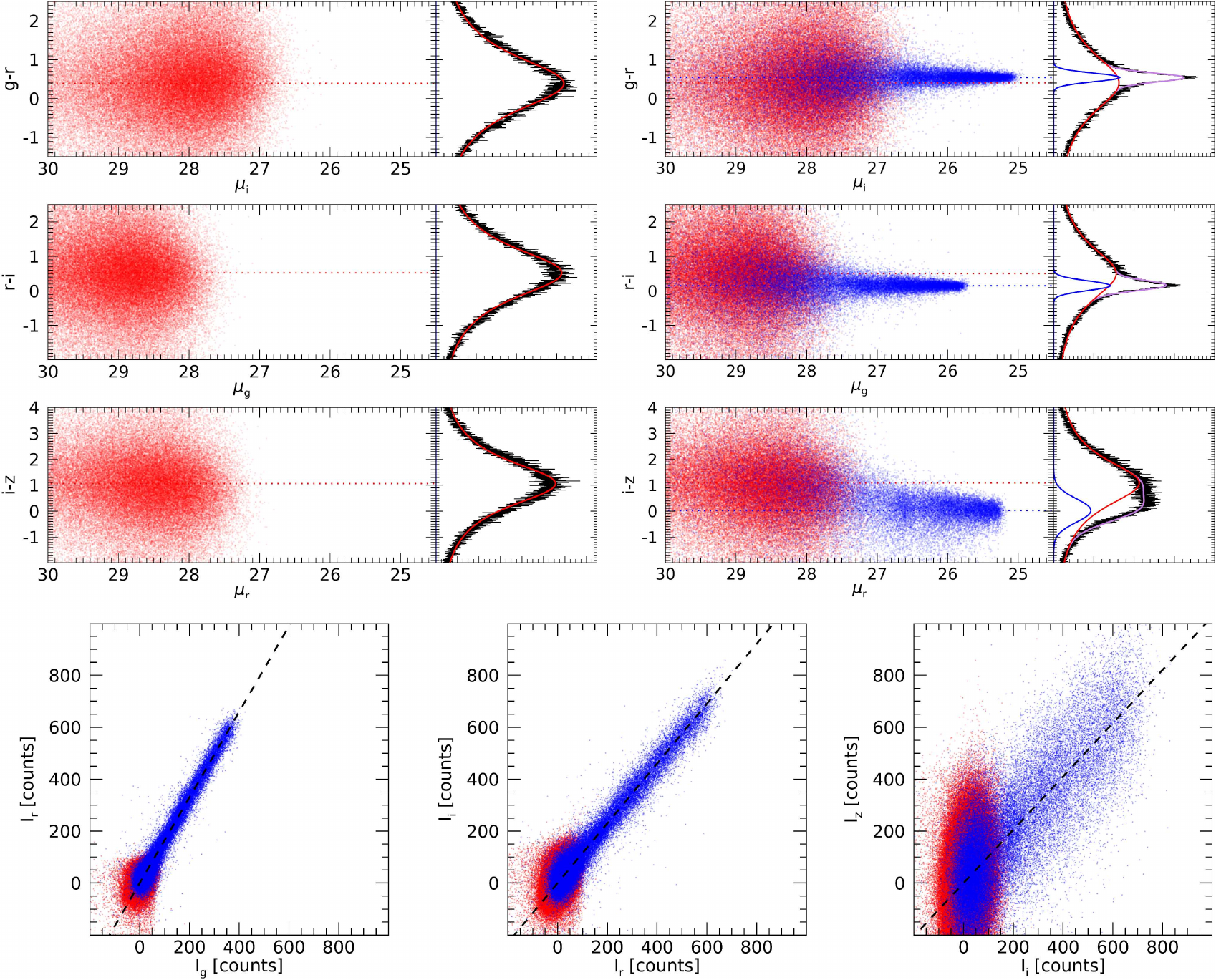}%0.635
    \caption{Color distribution of the noise in the Control Field and the resulting color distribution by adding an artificial constant color structure. {Upper} left panels: color vs. surface brightness values (red dots) of  pixels dominated by the noise. Their color distribution is very well fitted by a Lorentzian function (red curve) as shown in the right side of the plot. {Upper} right panel: same but adding a constant color structure (blue dots) fitted by a Gaussian function (blue curve). The purple curve is the result of combining a Lorentzian plus a Gaussian function. The horizontal dashed lines correspond to the calculated centers of the Lorentzian and Gaussian functions. {Bottom panels: Correlation between the flux of the different bands. The dotted line corresponds to the best fit using a linear least-squares approximation in one-dimension.}}

   \label{fig:Fit_noise}
\end{figure*}

In this Appendix, we expand on the details of the methodology used to obtain the colors of the Galactic cirri. We also perform a set of simulations to quantify the uncertainties in this measurement.

Due to the Poisson nature of the noise in the images, the color distribution of that noise (using two bands) follows a Lorentzian distribution. As the definition of magnitude involves the use of logarithm, only positive values of flux have physical sense. Therefore,  for a given pixel to enter in the color distribution, both bands A and B in that pixel must have a positive flux (for pixels with no source emission, only one in four pixels will enter in the distribution on average). Because in general, the depths of both bands are not the same (or widths of their Gaussian distributions), this causes the Lorentzian color distribution to be not centered in zero. In other words, given a pixel, its most likely color will be dominated by the flux of the shallower band (higher Gaussian dispersion). For example, two photometric bands A and B with limiting depths of $\mu_{A}$ = 25.5 and $\mu_{B}$ = 25.0 mag arcsec$^{-2}$ will produce a Lorentzian distribution centered in the color A-B = 0.5 mag arcsec$^{-2}$.

In the {upper} left panel of Fig. \ref{fig:Fit_noise}, we show the color distribution of pixels in the processed (see Sec. \ref{sec:proccesing}) Control field (see Fig \ref{fig:IR_opt}). Due to the absence of extended diffuse emission, the color distribution of the pixels in this field can be considered exclusively due to the noise of the image. We fit this distribution with a Lorentzian function. As can be seen, the fits show an excellent match with the observed color distribution. The centers of the fitted Lorentzian distributions are $x_{L}$ = 0.40, 0.50 and 1.14 mag for the \textit{g-r}, \textit{r-i} and \textit{i-z} colors respectively. While for the \textit{g-r} and \textit{r-i} colors the $x_{L}$ values match the depth differences of those bands, for \textit{i-z} this value differs by around 0.5 mag due to the higher sky fluctuations in the \textit{z} band. We note also, that the depth values provided in Section \ref{sec:Depth} are average values of the entire survey. The depths of a particular field may vary with respect to the average values of the whole survey due to the specific observational conditions in that field.

The color of any source present in the field will be added to this Lorentzian color distribution of the noise ({see upper right panels of Fig. \ref{fig:Fit_noise}}). There will be a transition from the color of the noise to the color of the source as the signal or surface brightness of the source increases.

In order to quantify the uncertainties in the photometry of the Cirri by the Lorentzian plus Gaussian methods, we perform a set of simulations injecting a structure of known color. This structure is used as mock Galactic cirri, and the process for recovering its color is done by the described methods. The injected structure has an area corresponding to approximately one fifth of the total area of the field and consists of a rectangle whose surface brightness decays on one axis, simulating the properties of the observed cirri, in which a wide range of surface brightness is found. The brightest region of this structure has a surface brightness of 25.5 mag arcsec$^{-2}$ and the faintest 30 mag arcsec$^{-2}$ in the \textit{g} band, well below the detection limits of the data. We show in the right panel of Fig. \ref{fig:Fit_noise} an example of the inclusion of such structure in the Control field. The blue points in the panels correspond to the pixels of the injected structure.

We perform the fitting of the color distribution of the field with the presence of this structure with a Lorentzian plus a Gaussian functions, as described in Section \ref{sec:colors}. {In the lower panels of Fig.\ref{fig:Fit_noise}, we plot the correlation between the fluxes in the different bands of the pixel distributions of noise + a constant color structure, corresponding to the flux fraction method applied in Section \ref{sec:Methodology}. It is worth noting the agreement showed by both methods in recovering the color of the structure.} This structure has colors of \textit{g-r} = 0.550, \textit{r-i} = 0.150 and \textit{i-z} = 0.050. We recovered the following colors \textit{g-r} = 0.542, \textit{r-i} = 0.130 and \textit{i-z} = 0.037, showing the accuracy and reliability of our proposed method. We note that although the color of the structure is constant, the recovered distribution is not a Delta function but a Gaussian function due to photometric uncertainties and the underlying sky background fluctuations.

\begin{figure}
  \centering
   \includegraphics[width=\columnwidth]{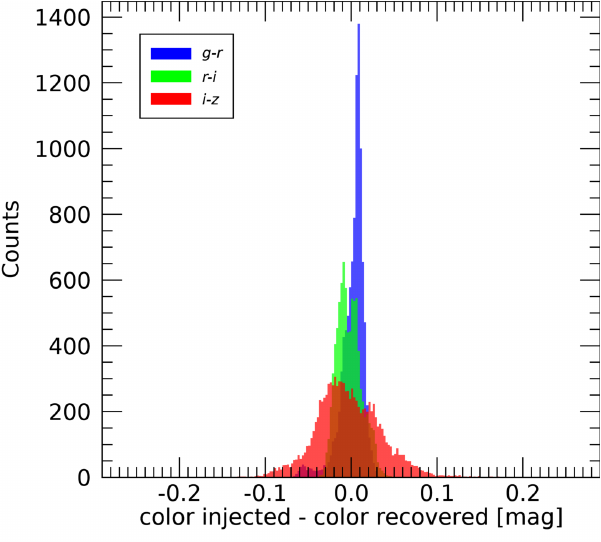}%0.635
    \caption{Differences between the injected and recovered colors of the mock Galactic cirri in the simulations.}

   \label{fig:Histo_injected}
\end{figure}

We performed a number of 10,000 injections and recoveries of such structures. In each iteration, we placed them in random positions within the field, avoiding a potential bias that could be introduced by locating the structures in the same area, for instance, anomalous sky fluctuation. The colors of the structures are also randomly drawn in each iteration, within the range [0,1] mag for \textit{g-r}, and [-0.5,0.5] mag for \textit{r-i} and \textit{i-z}. In Fig. \ref{fig:Histo_injected}, we plot the histogram of the difference between the injected and recovered \textit{g-r}, \textit{r-i} and \textit{i-z} colors. These colors follow an approximately Gaussian shape with slight deviations in their mean values with respect to 0 mag. The mean values of the distributions are -0.0066 mag, -0.0115 mag and -0.0085 mag for the \textit{g-r}, \textit{r-i} and \textit{i-z} colors respectively, and therefore, negligible. The 3$\sigma$ standard deviations of these distributions are: 0.04 mag, 0.04 mag and 0.11 mag for the \textit{g-r}, \textit{r-i} and \textit{i-z} colors respectively. Therefore, we assume these values as the uncertainties for the measurement of cirri colors by this method.

\section{Photometric maps of the Cirrus Field}\label{sec:AppendixD}
In Fig. \ref{fig:Filament_photC} and Fig. \ref{fig:Filament_photD}, we present the color maps of the Cirrus field discussed in Sec. \ref{sec:3}.

\begin{figure*}
  \centering
   \includegraphics[width=0.98\textwidth]{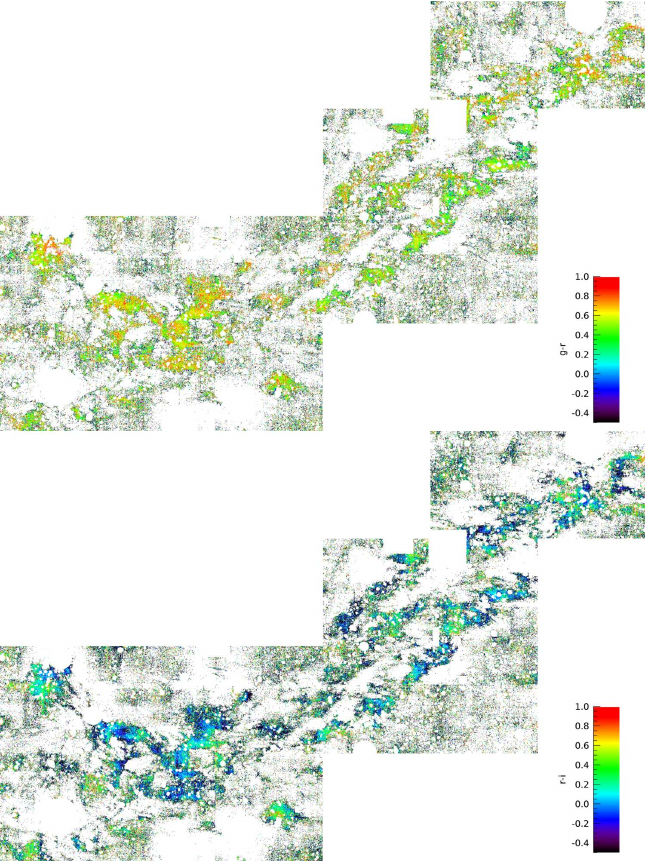}%0.635
   \caption{\textit{g-r} and \textit{r-i} color maps of the Cirrus Field.}
   \label{fig:Filament_photC}
\end{figure*}

\begin{figure*}
  \centering
   \includegraphics[width=0.98\textwidth]{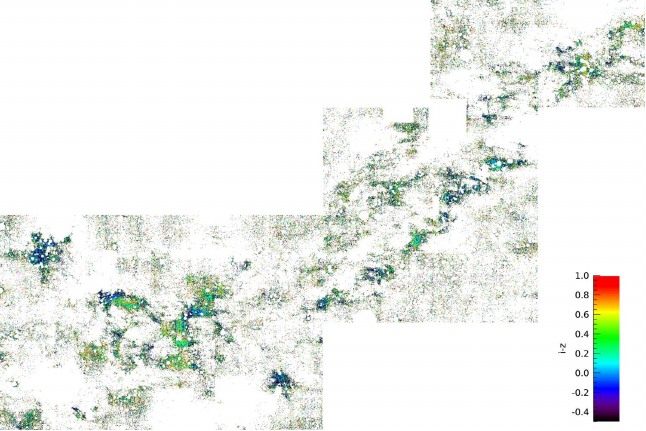}%0.635
   \caption{\textit{i-z} color map of the Cirrus Field.}
   \label{fig:Filament_photD}
\end{figure*}

\newpage

\section{Direct illumination of a star on Galactic dust}
\label{sec:nebula}

\begin{figure*}
  \centering
   \includegraphics[width=1.0\textwidth]{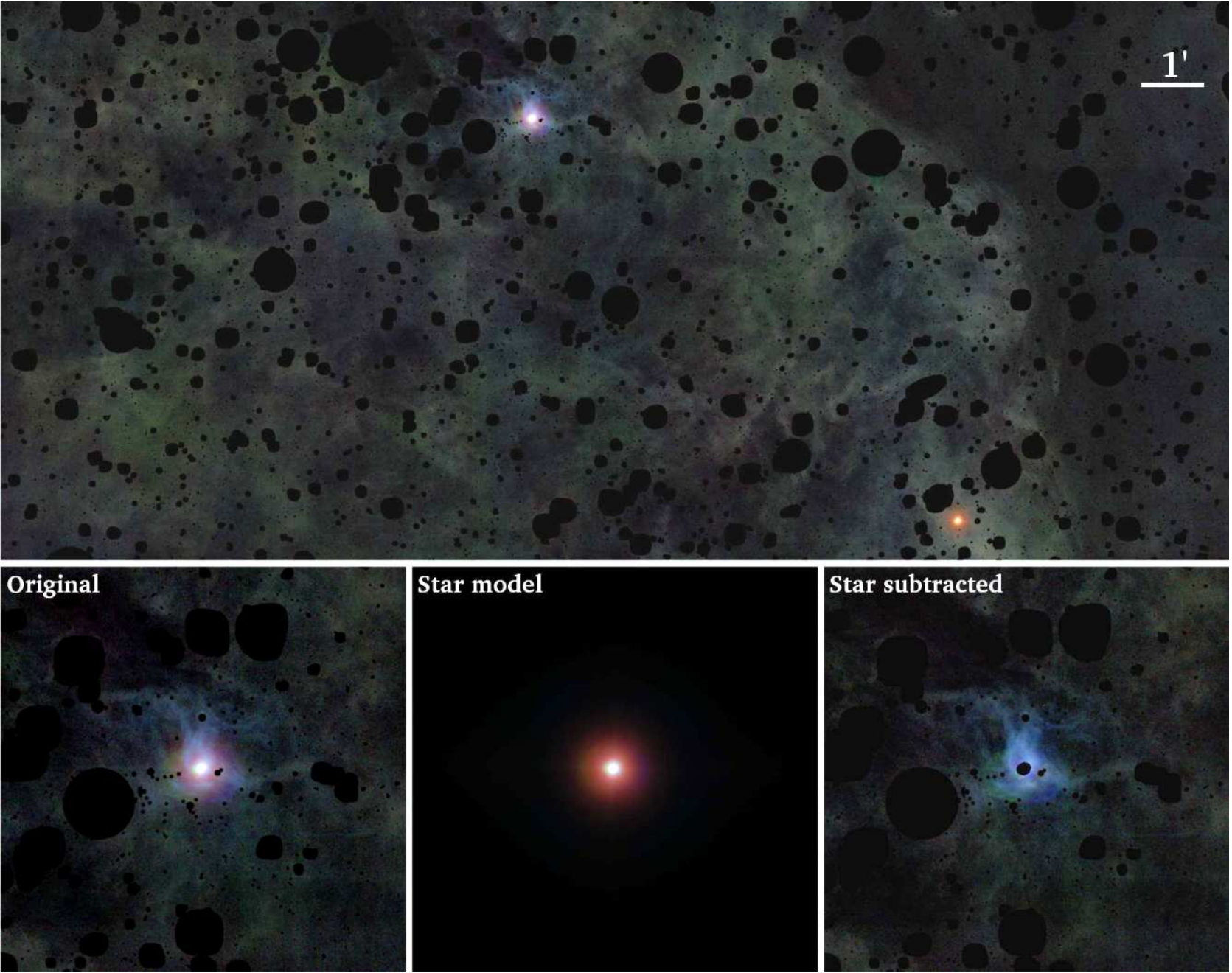}
    \caption{Composite color image using \textit{g} \textit{r} and \textit{i} SDSS bands of the region RA = [58.1,58.8], Dec = [-0.4, -0.1] at a pixel scale of 0.396 arcsec. All the sources have been masked and subtracted except the star that appears directly illuminating the dust screen and a random star in the bottom right side illustrating the appearance of all the stars without this fact. In the lower panels a zoom of 6.5$\times$6.5 arcmin of the area of interest is shown with the result of the subtraction of the star, showing more clearly the reflected area.}

   \label{fig:Nebula}
\end{figure*}

{The field depicted in Fig. \ref{fig:Nebula} contains a conspicuous cloud with a high far IR emission (high column density) adjacent to a fairly clean area (right part of the image) in which the sky can be calibrated accurately. The colors of the cirri in this area are compatible with the correlation shown in Fig.~\ref{fig:Corr_color} (S$_{100\mu m}$ = 14.1 MJy sr$^{-1}$, \textit{g-r} = 0.75, \textit{r-i} = 0.23, \textit{i-z} = 0.01). This field presents a very peculiar configuration. In the central upper part of the image there is what looks like a star in the vicinity of the dust screen, illuminating it directly. This low-brightness nebula was previously cataloged as a low-surface-brightness galaxy by \cite[][the \textit{0351-0019} galaxy]{1996ApJS..105..209I}, however, in our deeper data it seems to be a bright reflection on the dust by an adjacent star. To obtain the colors of this reflection, we subtract the star using the PSF in a similar way to that described in Sec. \ref{sec:proccesing}. Additionally, with this methodology, we are able to obtain the flux of the star, that being saturated, is the only way to obtaining it. The resulting colors of the star are \textit{g-r} = 0.51, \textit{r-i} = 0.21, \textit{i-z} = 0.20. To obtain the color of the nebula, we selected a small region, avoiding the central area where the inner internal reflections of the star could affect the obtained colors. As the area subtended by the nebula is small, we work at the original pixel scale of 0.396 arcsec of the image. This also prevent us of obtaining its colors in a similar fashion to that of the wide regions. At this spatial resolution, the signal to noise of the \textit{z} band is low, therefore we discard this band for this analysis. The colors of the nebula are \textit{g-r} $\approx$ 0.25 and \textit{r-i} $\approx$ 0.10. The fact that the nebula has colors considerably bluer than the adjacent star suggests a dust albedo that causes a bluing of the reflected light. Since this region is illuminated by a single star, and the colors of the star can be precisely known, this field is of potential interest for estimating the albedo of the dust in this region, maybe with a deeper and higher resolution dataset in the future.}

\end{appendix}

\end{document}